# Molecular theory of graphene chemical modification

**Elena F. Sheka**

Peoples' Friendship University of Russia,
sheka@icp.ac.ru

**Abstract**. The chapter is devoted to chemical uniqueness of graphene. In view of molecular theory, the latter is a consequence of benzenoid packing of carbon atoms, based on three neighbors to each carbon atom leaving the atom fourth valence electron (the odd one) on its own. Due to rather large C-C distances, the odd electrons of graphene molecules are correlated whereupon the molecules are radicalized and molecular chemistry of graphene is the chemistry of 'dangling bonds'. The chapter presents the description of the molecular-theory-dangling-bond view on graphene molecule hydrogenation and oxidation. Specific peculiarities of the graphene molecular chemistry related to size-dependent topochemical features are discussed, as well.

## 1. Introduction

The nickname of a 'miracle material' has been widely accepted by the graphene community. It normally means the material superior properties such as high electron mobility, high thermal conductivity, high value of the Young modulus, peculiar structure of its electron bands, etc. However, all these properties are only the outward manifestation of the graphene wonderful nature. The real miracle of graphene is that it is a union of two entities: the physical and chemical, each of which is unique in its own way. The above mentioned superior properties are characteristics of the graphene physical entity. Much less has been told about its chemical uniqueness. Moreover, physicists often refer to graphene as 'highly inert' material [1]. It is not the case. First of all, we must give an account of how much attractive to anyone the sound of the word "carbon". "Carbon" implies practically countless amount of scientific knowledge, both empirical and theoretical-computational. Obviously, this aspect has markedly contributed into unprecedented enthusiasm about graphene. However, the uniqueness of the graphene chemistry is generated, not by carbon atoms themselves but their packing in a flat honeycomb structure. The structure, based on benzenoid units, offers three neighbors to each carbon atom leaving the atom fourth valence electron on its own. These electrons form a pool of odd electrons whose behavior might be changed from the covalent bonding, characteristic for $\pi$ electrons, to free electrons of radicals when the interaction between the electrons becomes weaker and weaker. The two electron states belong to different limit cases in terms of the electron correlation: $\pi$ electrons are not correlated while radical electrons are strongly correlated so that two electrons with different spins occupy different places in the space. The odd electrons of benzenoid units of graphene in contrast with $\pi$-electrons of the benzene molecule are correlated, which is caused by the difference in the C-C bond lengths [2].

The odd electrons correlation and the extreme tight connection between the correlation degree and benzenoid bond structure make graphene material highly sensitive to any kind of external action such as morphological changing, chemical modification, mechanical loading and fixation, application of electric and magnetic field, and so forth thus making it structurally-and-electronically non stable. The "fluid" electronic structure [3] accompanied with the flexible space structure aggravated with equienergy topological phase transitions are the main reasons for failures of, say, stable working technologies for converting graphene from semimetal to semiconductor [4] as well as for postponing the application of the high-performance graphene up to 2035 or later [1]. Since until now chemical modification of graphene has been considered as the main way for achieving its best physical properties it is necessary to analyze all *pro* and *contra* on the way. Meeting the requirement, the current chapter is devoted to peculiarities of chemical modification of graphene in view of its chemical uniqueness. The chapter is organized as follows. Section 2 describes the grounds of computational chemistry of graphene on the basis of molecular theory. Section 3 concerns the graphene molecule hydrogenation. Oxygenation and reduction of the molecules are described in Section 4. Some peculiar effects caused by chemical modification are considered in Section 5. Section 6 summarized the most important points of the graphene molecular chemistry.

## 2. Grounds of Graphene Molecular Chemistry

Presented below is concentrated at the molecular essence of graphene considered from the viewpoint of the molecular theory of $sp^2$ nanocarbons. The theory is based on two main concepts, which involve 1) the odd-electron origin of the graphene electron system and 2) the correlation of the odd electrons. The latter turns out to play the governing role. As will be shown below, such an approach occurs very efficient in describing chemical properties of graphene.

### 2.1. Criteria of Odd Electrons Correlation

In spite of formally two-atomic unit cell of crystalline graphene, its properties are evidently governed by the behaviour of odd electrons of the hexagonal benzenoid units. The only thing that we know about the behaviour for sure is that the interaction between odd electrons is weak; nevertheless, how weak is it? Is it enough to provide a tight covalent pairing when two electrons with different spins occupy the same place in space or, oppositely, is it too weak for this and the two electrons are located in different places thus becoming spin correlated? This supremely influential molecular aspect of graphene can be visualized on the platform of the molecular quantum theory.

To exhibit a trend, a system computational experiment should be carried out meaning that a vast number of computations are to be performed as well as a great number of atoms are to be considered. At the same time, when speaking about electron correlation, one should address the problem to the configuration interaction (CI). However, neither full CI nor any of its truncated versions, clear and transparent conceptually, can be applied for the computational experiments, valuable for graphene nanoscience. Owing to this, techniques based on single determinants become the only alternative. Unrestricted Hartree-Fock (UHF) and unrestricted DFT (spin polarized, UDFT) approaches form the techniques ground and are both sensitive to the electron correlation, but differently due to different dependence of their algorithms on electron spins [5, 6]. The approach application raises two questions: 1) what are criteria that indicate the electron correlation in the studied system and 2) how much are the solutions of single-determinant approaches informative for a system of correlated electrons.

Answering the first question, three criteria, which highlight the electron correlation at the single-determinant level of theory, can be suggested [7]. This concerns the following characteristic parameters related to a benzenoid-packed molecule:

**Criterion 1: Misalignment of energy**

$$\Delta E^{RU} \geq 0 \text{ , where, } \Delta E^{RU} = E^R - E^U \text{ .}$$ )

Here, $E^R$ and $E^U$ are the total energies calculated by using the restricted and unrestricted versions of a software in use.

**Criterion 2: Misalignments of squared spin**

$$\Delta \hat{S}^2 \geq 0 \text{ ; } \Delta \hat{S}^2 = \hat{S}^2_U - S(S+1)$$ (2)

Here, $\hat{S}^2_U$ is the squared spin calculated within the applied unrestricted technique while $S(S+1)$ presents the exact value of $\hat{S}^2$.

**Criterion 3: Appearance of effectively unpaired electrons**

$$N_D \neq 0 \text{ .}$$

Here, $N_D$ is the total number of effectively unpaired electrons. The value is determined as

$$N_D = trD(r|r') \neq 0 \text{ and } N_D = \sum_A D_A \text{ .}$$ (3)

$D(r|r')$ [8] and $D_A$ [9] present the total and atom-fractioned spin density caused by the spin asymmetry due to the location of electrons with different spins in different spaces.

Criterion 1 follows from the well known fact that the electron correlation, if available, lowers the total energy [10]. Criterion 2 is the manifestation of the spin contamination of unrestricted single-determinant solutions [9, 11]; the stronger electron correlation, the bigger spin contamination of the studied spin state. Criterion 3 highlights the fact that the electron correlation is accompanied with the appearance of effectively unpaired electrons that provide the molecule radicalization [8, 9, 11].

Table 1 presents sets of the three parameters evaluated for a number of graphene molecules presented by rectangular $(n_a, n_z)$ fragments of graphene ($n_a$ and $n_z$ count the benzenoid units along armchair and zigzag edges of the fragment, respectively [12]), $(n_a, n_z)$ nanographenes (NGrs) below, by using the AM1 version of the semiempirical UHF approach implemented in the CLUSTER-Z1 codes [13]. To our knowledge, only this software allows for getting all the above three parameters within one computational session. Throughout the current chapter, the referred results are obtained in the framework of this approach.

As seen in the table, the parameters are certainly not zero, obviously greatly depending on the fragment size while their relative values are practically size-independent. The attention

Table 1. Identifying parameters of the odd electron correlation in the rectangular graphene molecules [7]

| $(n_a, n_z)$ nanographenes | Odd electrons $N_{odd}$ | $\Delta E^{RU*}$ kcal/mol | $\delta E^{RU}$ %** | $N_D$, e⁻ | $\delta N_D$, %** | $\Delta\hat{S}_U^2$ |
|---|---|---|---|---|---|---|
| (5, 5) | 88 | 307 | 17 | 31 | 35 | 15.5 |
| (7, 7) | 150 | 376 | 15 | 52.6 | 35 | 26.3 |
| (9, 9) | 228 | 641 | 19 | 76.2 | 35 | 38.1 |
| (11, 10) | 296 | 760 | 19 | 94.5 | 32 | 47.24 |
| (11, 12) | 346 | 901 | 20 | 107.4 | 31 | 53.7 |
| (15, 12) | 456 | 1038 | 19 | 139 | 31 | 69.5 |

* AM1 version of UHF codes of CLUSTER-Z1 [13]. Presented energy values are rounded off to an integer

* The percentage values are related to $\delta E^{RU} = \Delta E^{RU} / E^R(0)$ and $\delta N_D = N_D / N_{odd}$, respectively

should be called to rather large $N_D$ values, both absolute and relative, indicating a considerable radicalization of the studied nanographenes. It should be added as well that the relation $N_D = 2\Delta\hat{S}_U^2$, which is characteristic for spin contaminated solutions in the singlet state [9], is rigidly kept over all the fragments. Therefore, the odd electrons correlation for the studied molecules is quite significant. Generally, the conclusion can be addressed to nanographenes of any shape. However, the numeric identifying parameters will obviously depend on the species shape due to a particular role of circumference atoms, which will be shown below. Nevertheless, it is possible to state that any nanographene– i.e. graphene molecule – is considerably radicalized, non depending on its size and shape.

## 2.2. Quantitative Description of Graphene Molecule Radicalization

The UHF computational scheme leads to spin-contaminated solutions due to which there is a question, mentioned in the previous section: how reliable and informative are the obtained solutions? Thoroughly analyzed in [14], the following answers to the questions have been obtained: 1) the broken symmetry UHF approach allows obtaining the exact energy of pure-spin states; 2) the approach provides exact determination of the magnetic constant; 3) the approach forms the grounds for computational chemistry of electron-correlated graphene molecules.

Criterion 3 lays the foundation for the basic concepts of the computational chemistry of graphene. Firstly shown by Takatsuka, Fueno, and Yamaguchi [8], the correlation of weakly interacting electrons is manifested through the density matrix $D(r|r')$, named as the distribution of 'odd' electrons. The function was proven to be a suitable tool to describe the spatial separation of electrons with opposite spins, and its trace $N_D = trD(r|r')$ was interpreted as the total number of these electrons [8, 15]. The authors suggested $N_D$ to manifest the radical character of the species under investigation. Over twenty years later, Staroverov and Davidson changed the term by the 'distribution of *effectively unpaired electrons*' [9, 16] emphasizing that not all the odd electrons may be taken off the covalent bonding. Even Takatsuka et al. mentioned [8] that the function $D(r|r')$ can be subjected to the population analysis within the framework of the Mulliken partitioning scheme. In the case of a single Slater determinant, $N_D$ can be determined as [9]

$$N_D = trDS ,$$ (4)

where,

$$DS = 2PS - (PS)^2 .$$ (5)

Here, $D$ is the spin density matrix $D = P^\alpha - P^\beta$ while $P = P^\alpha + P^\beta$ is a standard density matrix in the atomic orbital basis, and $S$ is the orbital overlap matrix ($\alpha$ and $\beta$ mark different spins). The population of effectively unpaired electrons on atom $A$ is obtained by partitioning the diagonal of the matrix $DS$ as

$$N_{DA} = D_A = \sum_{\mu \in A} (DS)_{\mu\mu} ,$$ (6)

so that

$$N_D = \sum_A N_{DA} .$$ (7)

Staroverov and Davidson showed [9] that the atomic population $N_{DA}$ is close to the Mayer free valence index $F_A$ in a general case while in the singlet state $N_{DA}$ and $F_A$ are identical. Thus, plotting $N_{DA}$ over atoms gives a visual picture of the actual radical electrons distribution that, in its turn, exhibits atoms with the enhanced chemical reactivity.

The UHF computational algorithm realized in the *NDDO* approximation [17] (the basis for the AM1/PM3 semiempirical techniques in CLUSTER-Z1) suggests the exact determination of both $N_D$ and $N_{DA}$ as follows [18]

$$N_D = \sum_A N_{DA} = \sum_{A=1}^{NAT} \sum_{i \in A}^{NORBS} \sum_{B=1}^{NAT} \sum_{j \in B}^{NORBS} D_{ij} .$$ (8)

Here, $D_{ij}$ are elements of the spin density matrix $D$, *NORBS* and *NAT* mark the number of orbitals and atoms, respectively. The expression can be easily implemented into UHF codes, as was done for CLUSTER-Z1, thus opening a possibility to quantitatively describe the extent of radicalization of graphene molecules and thereby predetermining their chemical behavior, which will be shown in the next Sections. Before going to these topics, turn our attention to the cause of the correlation of odd electrons in graphene.

### 2.3. The Cause of Odd Electrons Correlation in sp² Nanocarbons

The odd electrons story of nanocarbons is counted from the discovery of the benzene molecule by Michael Faraday in 1825. However, only a hundred years later Hückel suggested the explanation of the deficiency of hydrogen atoms in the molecule to complete the valence ability of its carbon atoms. Extra, or odd, electrons were named as π electrons that, in contrast to trigonal σ electrons, interact much weaker while providing the additional covalent coupling between neighboring atoms. The two electrons are located in the same space, and their spins are subordinated to the Pauli law. Formally, this π electrons view, which lays the foundation of the aromaticity concept, has been expanded over all $sp^2$ nanocarbons and has been shared by a number of material scientists in the field until now. However, the concept does not take into account a crucial role of the distance between neighboring odd electrons. As seen in Fig. 1, which presents a plotting of the total number of effectively unpaired electrons $N_D$ as a function of the C-C distance in the ethylene molecule, the bond stretching from its equilibrium value of 1.326Å up to $R_{crit} = R_{cov}^{C-C} = 1.395$Å does not cause the appearance of the unpaired electrons so that the relevant π electrons are fully covalently bound. However, above $R_{crit}$ the number $N_D$ gradually increases up to a clearly vivid knee that is characterized by $N_D \cong 2$ at $R=1.76$Å, which evidences a complete radicalization of the previous π electrons. Further stretching concerns mainly two σ electrons that, once fully covalently bound until $R=1.76$Å, gradually become unpaired just repeating the fortune of π electrons resulting in $N_D \cong 4$ at 2.5Å. In the figure, two vertical arrows circumscribe the region of the C-C bond lengths typical for fullerenes, carbon nanotubes, and graphene. As follows from this picture, the radicalization of π electrons of these molecules is still nearly at the beginning. Nevertheless, the number of unpaired electrons becomes quite significant to determine specific molecule chemistry.

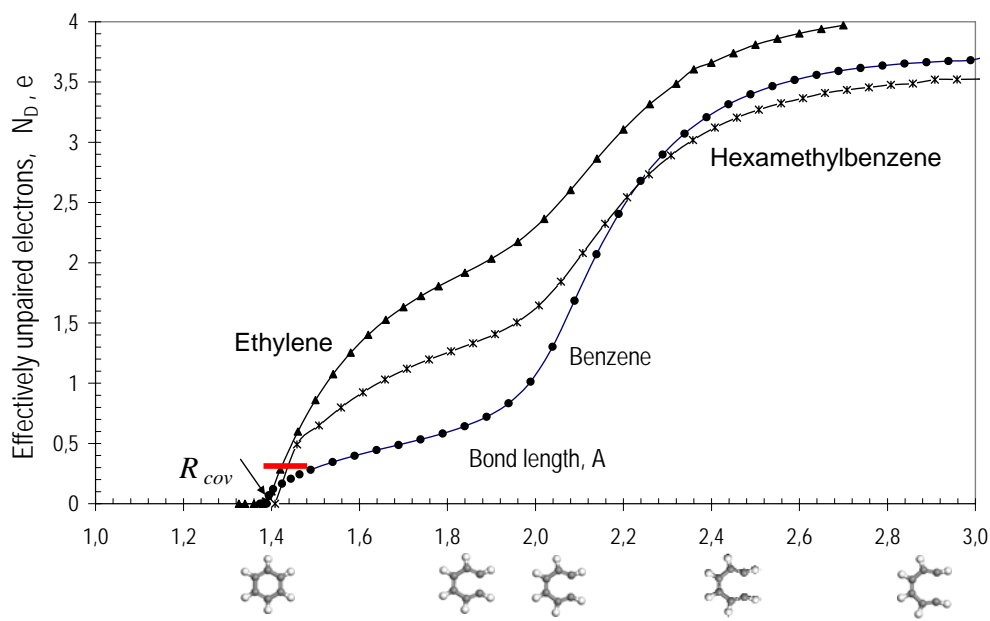

**Figure 1**. The total number of effectively unpaired electrons $N_D$ accompanying a stretching of C-C bond in ethylene. $R_{cov}^{C-C}$ marks the extreme distance that corresponds to the completion of the covalent bonding. $R_{rad}^{C-C}$ matches a completion of homolytic bond cleavage. Red horizontal bar marks the interval of the C-C bond lengths characteristic for $sp^2$ nanocarbons.

In spite of clear explanation where unpaired electrons are coming from, the question about their existence still remains due to suspicion of their attribution to an *artifact* caused by the limitations of the single-determinant calculations. However, recent examinations have convincingly shown that unpaired electrons are a physical reality [19] (see comments in [14]).

### 3. Molecular Theory of Graphene Hydrogenation

According to UHF calculations, C-C bond length of graphene molecules fill an interval (It should be mentioned that the application of the restricted version of the same program results in practically non-dispersive value of the C-C bond length of 1.42Å.) that depends on the molecule size. Thus, for the (5, 5) NGr the interval constitutes 1.322-1.462 Å while for (11, 11) NGr it is 1.284-1.468 Å. The relative number of bonds whose length exceeds $R_{crit}$ is 62% and 81% in the two cases, respectively. This leads to a considerable amount of the effectively unpaired electrons, total numbers of which are listed in Table 1 for different nanographenes. From the chemical viewpoint, $N_D$ values describe the molecules chemical susceptibility [2] while $N_{DA}$ presents atomic chemical susceptibility (ACS) of atom A. These two quantities become the main parameters of the computational chemistry of graphene molecules.

Figure 2 exhibits a typical ACS image map presented by the $N_{DA}$ distribution over atoms of the (5, 5) NGr. The molecule edges are not terminated, and the $N_{DA}$ image map has a characteristic view with a distinct framing by edge atoms since the main part of the unpaired electrons is concentrated in this area. This map presents a typical chemical portrait of any graphene molecule and exhibits the exceptional role of the circumference area, which looks like a typical 'dangling bonds' icon [20]. At the same time, the ACS image map intensity in the basal plane is of ~0.3 $e$ in average so that the basal plane should not be discounted when it comes to chemical modification of the molecule thus disproving the common idea of its chemical inertness. Moreover, basing on the $N_{DA}$ value and choosing the largest of them as a quantitative pointer of the target atom for the coming chemical attack, one can suggest the algorithmic 'computational syntheses' of the molecule polyderivatives [21].

The absolute $N_{DA}$ values of the (5, 5) NGr, shown in Fig. 2c according to the atom numbering in the output file, clearly exhibit 22 edge atoms involving 2x5 *zg* and 2x6 *ach* ones have the highest $N_{DA}$ thus marking the perimeter as the most active chemical space of the molecule. The molecule hydrogenation starts on atom 14 (star-marked in Fig. 2c) according to the largest $N_{DA}$ in the output file. The next step of the reaction involves the atom from the edge set as well, and this is continuing until all the edge atoms are saturated by a pair of hydrogen atoms each since all 44 steps are accompanied with the high-rank $N_{DA}$ list where edge atoms take the first place [22]. Thus obtained hydrogen-framed (5, 5) NGr molecule is shown in Fig.3 alongside with the corresponding ACS image map. Two equilibrium structures are presented. The structure in panel *a* corresponds to the optimization of the molecule structure without any restriction. In the second case, positions of the edge carbon atoms and framing hydrogen atoms under optimization were fixed. In what follows, we shall refer to the two structures as a free standing and fixed membrane, respectively.

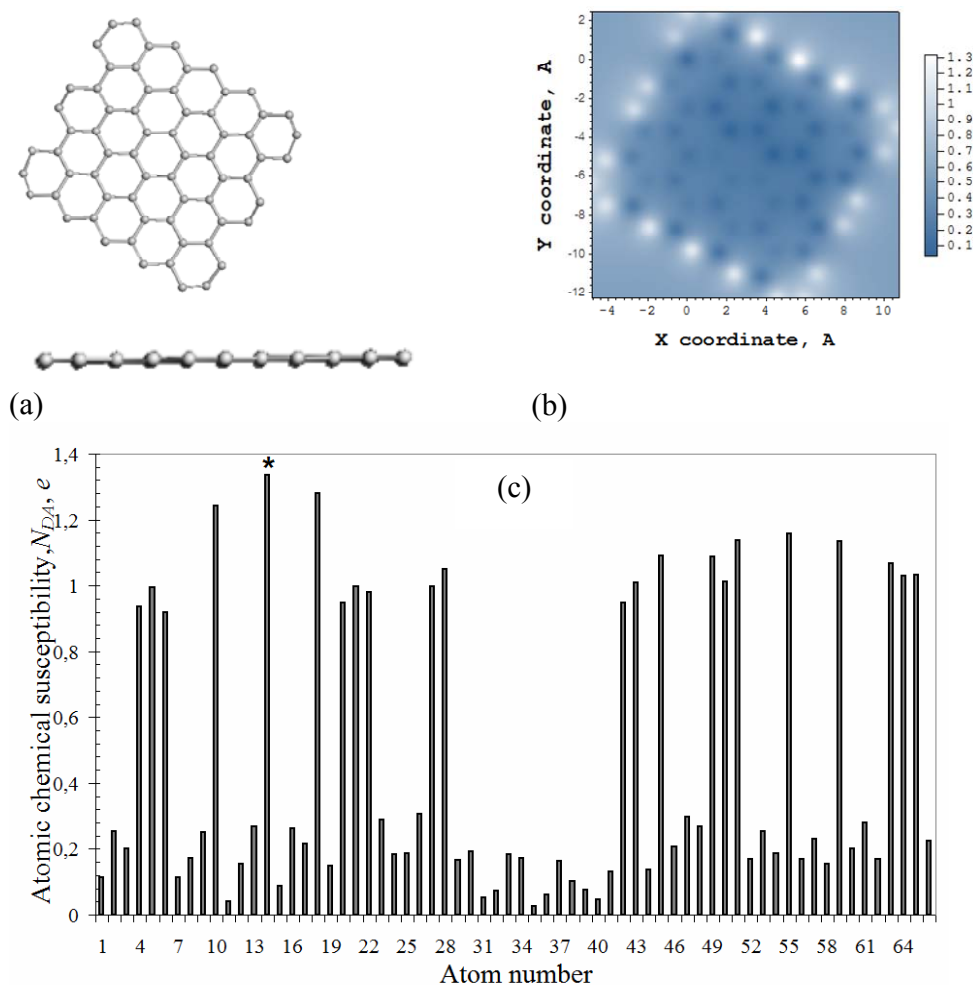

(a)

(b)

(c)

**Figure 2**. Top and side views of the equilibrium structure of (5,5) NGr molecule (a); $N_{DA}$ image map (b) and $N_{DA}$ distribution over atoms according to atom numbers in the output file (c) [22].

The chemical portraits of the structures shown in Fig.3b and Fig.3d are quite similar and reveal the transformation of brightly shining edge atoms in Fig.2b into dark spots. The addition of two hydrogen atoms to each of the edge ones saturates the valence of the latter completely, which results in zeroing $N_{DA}$ values as is clearly seen in Fig.3e. Chemical activity is shifted to the neighboring basal atoms and retains higher in the vicinity of $zg$ edges, however, differently in the two cases. The difference is caused by the redistribution of the C-C bond lengths of the free standing membrane when it is fixed over perimeter, thus providing different starting conditions for the hydrogenation of the two membranes.

Besides the two types of initial membranes, the hydrogenation depends on other factors, such as 1) the hydrogen species in use and 2) the accessibility of the membrane sides to the hydrogen. Even these circumstances evidence the hydrogenation of graphene to be a complicated chemical event that strongly depends on the initial conditions, once divided into 8 adsorption modes in regard to atomic or molecular adsorption; one- or two-side accessibility of membranes; and free or fixed state of the membranes perimeter. Only two ones of the latter correspond to the experimental observation of hydrogenated specimens discussed in [23], namely: two-side and one-side atomic hydrogen adsorption on the fixed membrane. Stepwise hydrogenation of the (5, 5) NGr molecule was considered in details in [22].

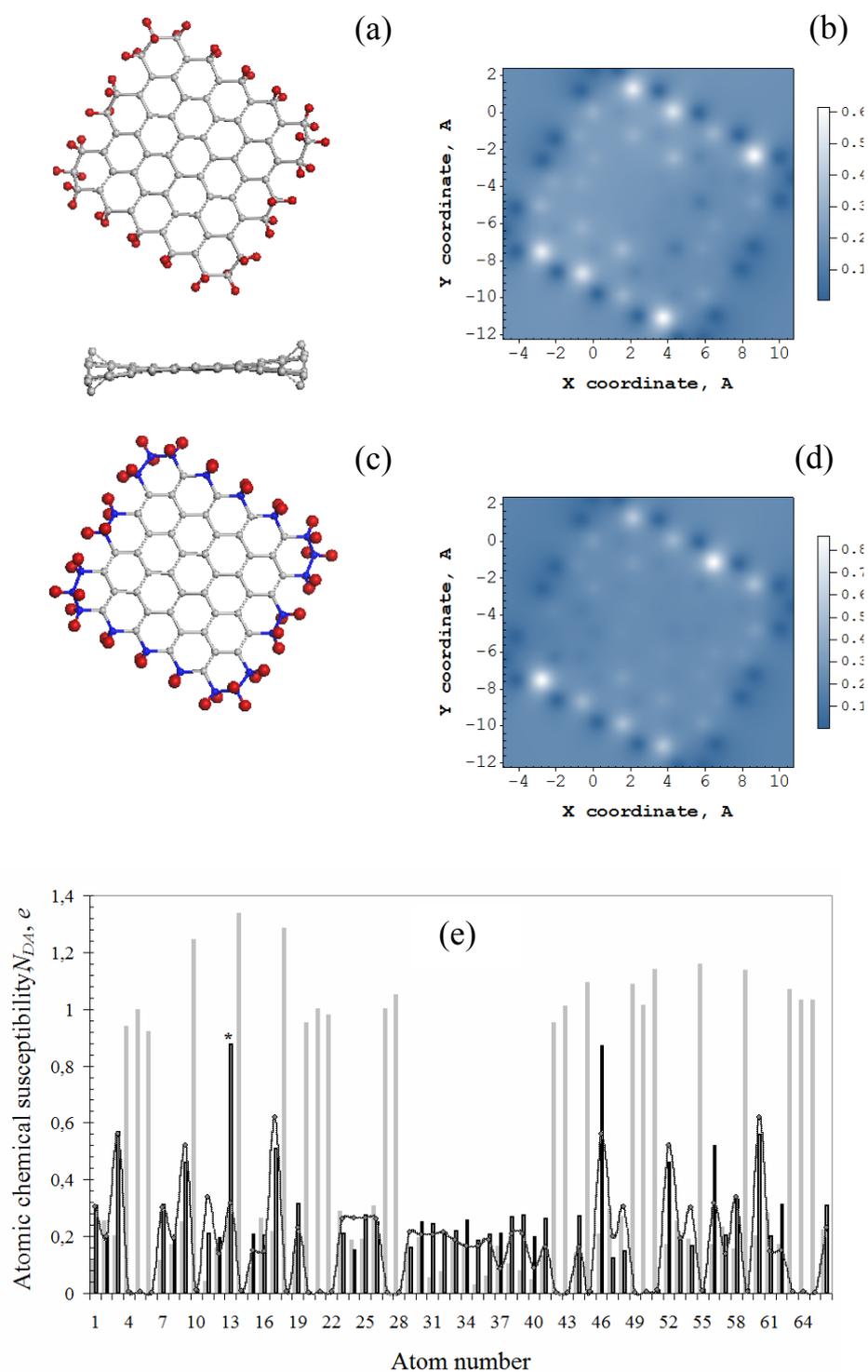

**Figure 3**. Equilibrium structures of free standing (a) and fixed (c) (5,5) NGr membrane; $N_{DA}$ image maps (b, d) and $N_{DA}$ distribution over atoms according to atom numbers in the output file (e) [22]. Gray and red balls mark carbon and hydrogen atoms, respectively. Light gray histogram plots ACS data for the pristine (5,5) NGr molecule. Curve and black histogram are related to membranes in panels *a* and *c*, respectively.

### 3.1. Two-Side Atomic Adsorption of Hydrogen on Fixed Membrane

 The hydrogenation concerns the basal plane of the fixed hydrogen-framed membrane shown in Fig.3c that is accessible to hydrogen atoms from both sides. As seen in Fig.3e, the first hydrogenation step occurs on basal atom 13 marked by a star. Since the membrane is accessible to hydrogen from both sides, a choice in favor of deposition of the hydrogen atom above the carbon plane ('up') was done following the lowest-total-energy (LTE) criterion.

After the first deposition, the $N_{DA}$ map revealed carbon atom 46 for the next deposition (see H1 $N_{DA}$ map in Fig.4). The LTE criterion favors the 'down' position of structure H2 shown in the figure. The second deposition highlighted next targeting carbon atom 3 (see $N_{DA}$ map of H2 hydride), the third adsorbed hydrogen atom activated target atom 60, the fourth did the same for atom 17, and so forth. Checking up and down depositions in view of the LTE criterion, a choice of the best configuration was performed and the corresponding equilibrium structures for a selected set of the (5, 5) NGr polyhydrides from H1 to H11 are shown in Fig.4. A set of the next structures from the 15[th] step up to the final 44[th] step is presented in Fig.5. A random distribution of the adsorbed hydrogen atoms, presented by the set, well correlates with experimentally observed picture [24]. The set is completed by regular structure at the 44[th] step. This structure H44, including framing hydrogen atoms, is shown at the bottom of the figure and presents a computationally synthesized fully saturated chairlike (5, 5) nanographane (NGra) that is in full accordance with the experimental observation of the graphane crystalline structure [23]. A complete set of the two-side-adsorption products which involves both a free-standing and fixed membranes forms polyhydrides family 1 shown in Fig. 33.6.

### 3.2. One-Side Atomic Adsorption of Hydrogen on Fixed Membrane

Starting from the first step of the hydrogenation described in the previous section, the reaction proceeded further with the second and all the next steps of the up deposition only. As previously, the choice of the target atom at each step was governed by the high-rank $N_{DA}$ values. Figure 6c presents the saturated graphene polyhydride related to the final 44[th] step over the basal atoms. A peculiar canopy shape of the carbon skeleton of the hydride is solely provided by the formation of the table-like cyclohexanoid units. However, the unit packing is quasi-regular which may explain the amorphous character of the polyhydrides formed at the outer surface of graphene ripples observed experimentally [23]. A complete set of the one-side-adsorption products which involves both a free-standing and fixed membranes forms polyhydrides family 2 shown in Fig. 6.

As for the hydrogen coverage, Fig. 7 presents the distribution of C-H bond lengths of the two discussed saturated graphene polyhydrides. In both cases, the distribution consists of two parts, the first of which covers 44 C-H bonds formed at the molecule skeleton edges.

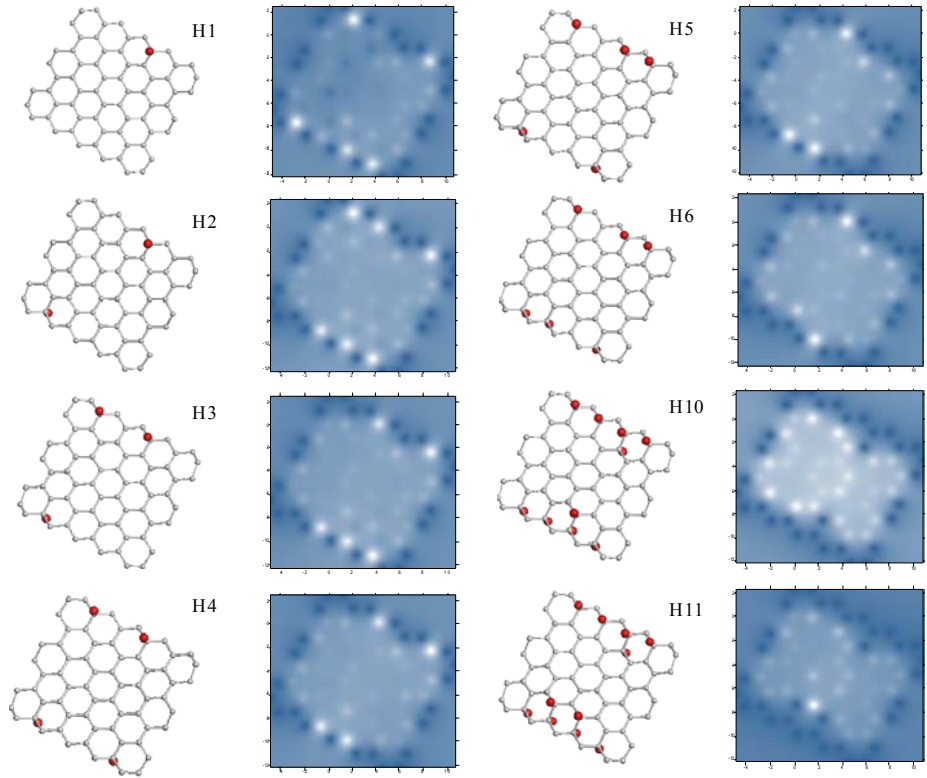

**Figure 4**. Equilibrium structures (left) and $N_{DA}$ image maps (right) of graphene hydrides related to initial stage of the basal-plane hydrogenation. HKs denote hydrides with K hydrogen atoms deposited on the membrane basal plane [22]. Framing hydrogen atoms are not shown to simplify the structure image presentation. Atoms marking see in caption to Fig. 3.

Obviously, this part is identical for both hydrides since the bonds are related to the framing atoms. The second part covers C-H bonds formed by the hydrogen atoms attached to the basal plane. As seen in the figure, in the case of polyhydride 1, C-H bonds are practically identical with the average length of 1.126Å and only slightly deviate from those related to framing atoms. This is just a reflection of the regular graphane structure of the hydride H44 shown in Fig.5 similarly to highly symmetric fullerene hydride $C_{60}H_{60}$ [25]. In contrast, C-H bonds on a canopy-like carbon skeleton of hydride 2 are much longer than those in the framing zone, significantly oscillate around the average value of 1.180Å. In spite of the values markedly exceed a 'standard' C-H bond length of 1.11Å, typical for benzene, those are still among the chemical C-H bonds, whilst stretched, since the C-H bond rupture occurs at the C-H distance of 1.72Å [26]. A remarkable stretching of the bonds points to a considerable weakening of the C-H interaction for polyhydride 2 in comparison with polyhydride 1, which is supported by the energetic characteristics of the hydrides, as well [22]. The total energies of both hydrides are negative by sign and gradually increase by the absolute value when the number of adsorbed atoms increases. However, the absolute value growth related to polyhydrides 2 is slowing down starting at step 11 in contrast to the continuing growth for polyhydrides 1 [22]. This retardation obviously shows that the one-side addition of hydrogen to the fixed membrane of polyhydrides 2 at the coverage higher than 30% is more difficult than in the case of the two-side addition of polyhydrides 1, for which the reaction of the chemical attachment of the hydrogen atoms is thermodynamically profitable through over the covering up to the 100% limit. In contrast, the large coverage for polyhydrides 2 becomes less and less profitable so that at final steps the hydrogen adsorption and desorption become competitive.

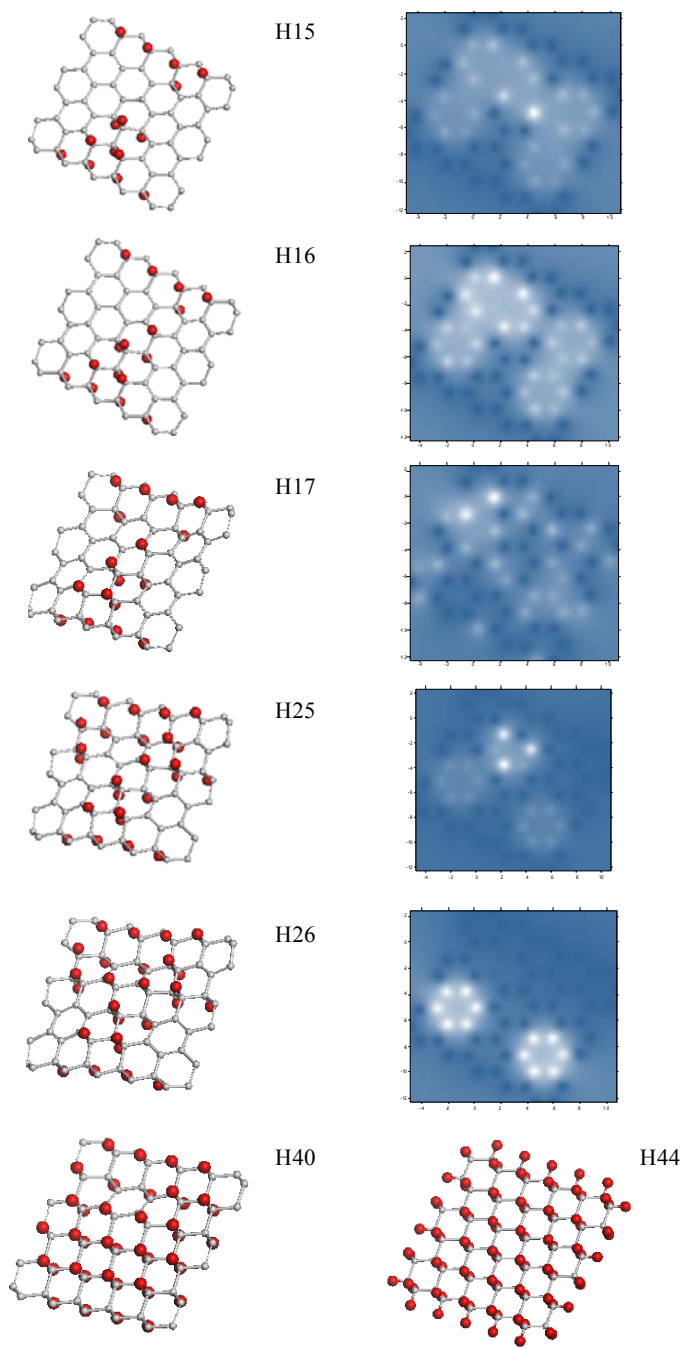

**Figure 5**. Equilibrium structures (left) and $N_{DA}$ image maps (right) of graphene hydrides from H15 to H44 [22]. Framing hydrogen atoms are shown for H44 only. Atoms marking see in caption to Fig. 3.

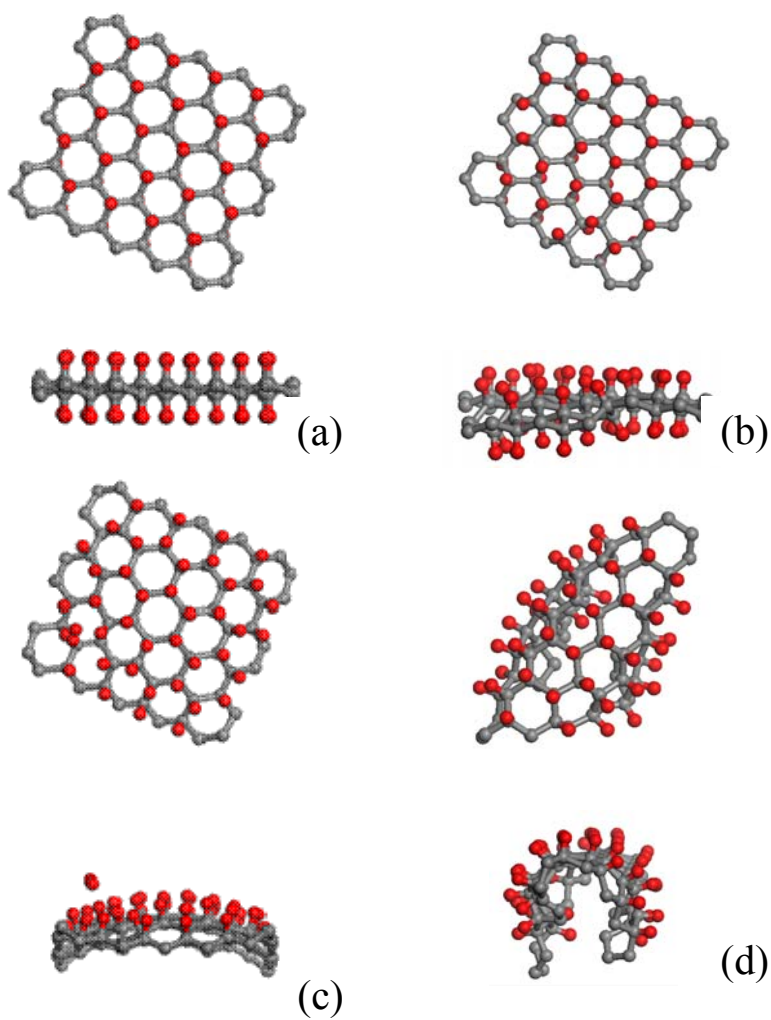

**Figure 6**. Equilibrium structures (top and side views) of fixed (a, c) and free standing (b, d) (5, 5) NGr membranes under two-side (a, b) (polyhydrides I) and one-side (c, d) (polyhydrides II) hydrogen adsorption. Framing hydrogen atoms are not shown to simplify the structure image presentation. Atoms marking see in caption to Fig.3.

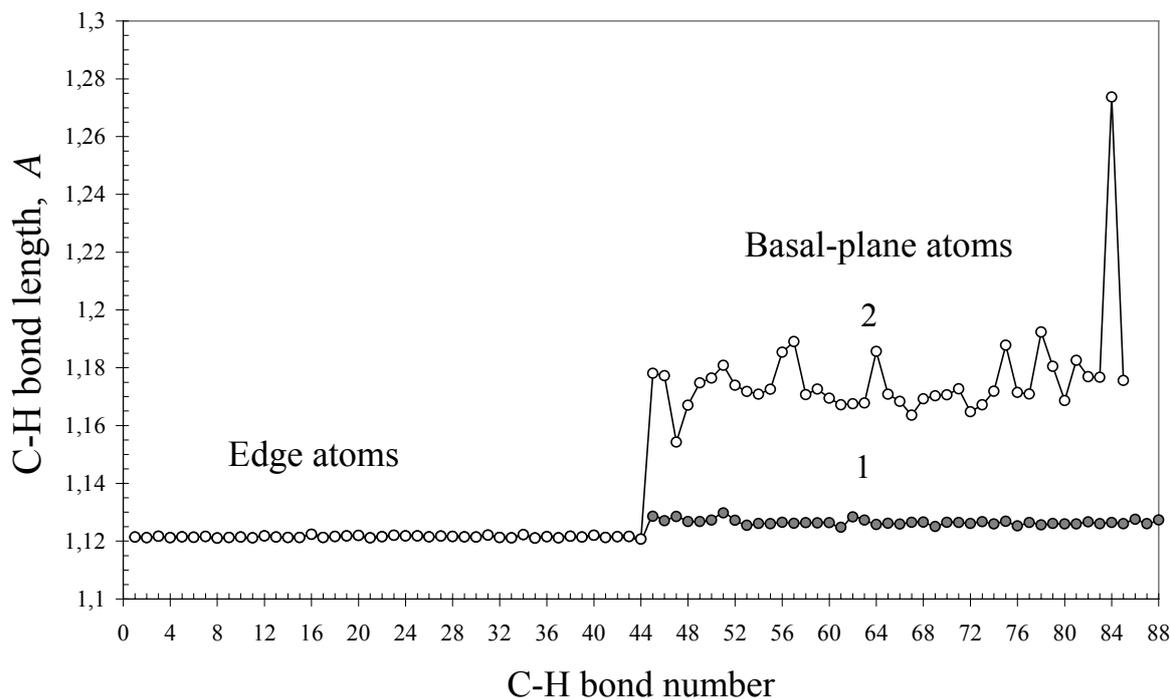

**Figure 7.** Equilibrium structures of (5,5) NGr polyhydrides and polyoxides related to the 1st, 2nd, 3rd, 4th, 5th, and 22nd (21st in the case of OH-framing, see text) steps of first-stage reactions. Roman figures mark the obtained GOs. Red and blue balls mark hydrogen and oxygen atoms, respectively [28].

### 3. Molecular Theory of Graphene Oxydation

In view of the molecular theory, the hot points of the GO chemistry can be presented as the following questions:

• What is the role of individual oxidants, such as O, OH, and COOH, in the oxidation of a graphene molecule?

• How does the coupling of the oxidants with the molecule depend on the local place of the addition?

• How is the highest degree of the derivatization under graphene molecule oxidation?

• What is the structure of the carbon skeleton of GO and can be the regular structure achieved?

• What can chemical composition of GO be obtained under conditions close to experimental ones?

Answering each of the questions presents a topic for a valuable computational study. A temptation to simultaneously answer all the questions raises a problem of a system approach, or, by other words, of an extended computational experiment. The computational molecular theory of graphene is the very feasible method in the framework of which the approach can be realized. The latter means that not selected individual computations but a solution of a number of particularly arranged computational problems is the aim of the computational experiment. Evidently, the bigger set, the more colors participate in drawing the image of the GO chemistry.

The first attempt of such system computational approach including about 400 computational jobs has been recently undertaken [27, 28]. The pristine (5, 5) NGr molecule was the basis of the performed computational experiment.

Stepwise oxidation of the (5, 5) NGr molecule in the form of free standing membrane was considered similarly to the hydrogenation described above. On the background of a tight similarity in both processes, in general, important difference of the events concerns the fact that instead of atomic hydrogens, which were attacking agents in the first case, a set of oxidants consisting of oxygen atoms O, hydroxyls OH, and carboxyls COOH had to be considered in the latter case. A detailed description of the molecule oxidation is given in [27, 28]. A comparable study of the results related to different configurations of GOs has led the foundation for the suggested conclusions. The oxidation was considered as a two-stage reaction, the first of which covered first 22 steps that involve edge atoms of the molecule only, on one hand, and all the oxidants, on the other. The second stage combines results related to both edge and basal atoms for oxidation by OH and COOH and basal atoms only when attaching atomic oxygen.

### 4.1. The First-Stage Oxidation of the (5, 5) NGr Molecule

#### 4.1.1. Homo-Oxidant Action

Figure 8 presents equilibrium structures of the O-, OH-, and COOH- multi-fold GOs at the first five and the 22nd steps of oxidation. For comparison, a similar first stage of the (5, 5) NGr molecule hydrogenation is added. In all the cases, each next step of a subsequent addition was governed by the highest rank $N_{DA}$ value at a preceding step. As seen in the figure, the difference in the starting points at the second step has actually led to the different progress of the hydrogenation and oxidation processes in space.

In the case of O-GOs, the obtained 22-fold GO (GO I) presents a completed framing of the molecule. In the case of OH-GOs, a successive OH- framing continued up to the 12th step. After optimization of the 13-fold OH-GO, its equilibrium structure revealed the dissociation of one previously attached OH group and the formation of C-H and C=O bonds instead. All the next steps of oxidation up to the 21st one did not cause any similar transformation and the final GO II shown in the last row in the figure was obtained. The molecule became not flat practically since the third step. Next steps increased the corrugation, but the skeleton deformation became more smoothed by the end of framing. In the case of COOH-oxides, sterical constraints additionally complicated framing as seen in the case of the 22-fold COOH-framed GO III shown at the bottom of Fig.8.

The first-stage framing considerably suppresses the chemical activity of the GO molecule edge atoms, but, evidently, not to the end. Thus, carbonyls, formed over GO I, provide practically full suppression of the chemical activity of the edge atoms. The further oxidation of the molecule, starting at the 23rd step, occurs at the basal plane. In contrast, the chemical activity at the circumference of GO II and GO III is not fully suppressed that is why both edge and basal atoms should be taken into account when considering further oxidation.

The analysis of energetic regularities completes the consideration of homo-oxidant framing of the (5, 5) NGr molecule. Figure 9a presents a set of plottings related to per step coupling energies (PCEs). At each step, the PCE was determined as

$$E_{cpl}^{pst}(GO_n) = \Delta H(GO_n) - \Delta H(GO_{n-1}) - \Delta H(Oxd). \tag{9}$$

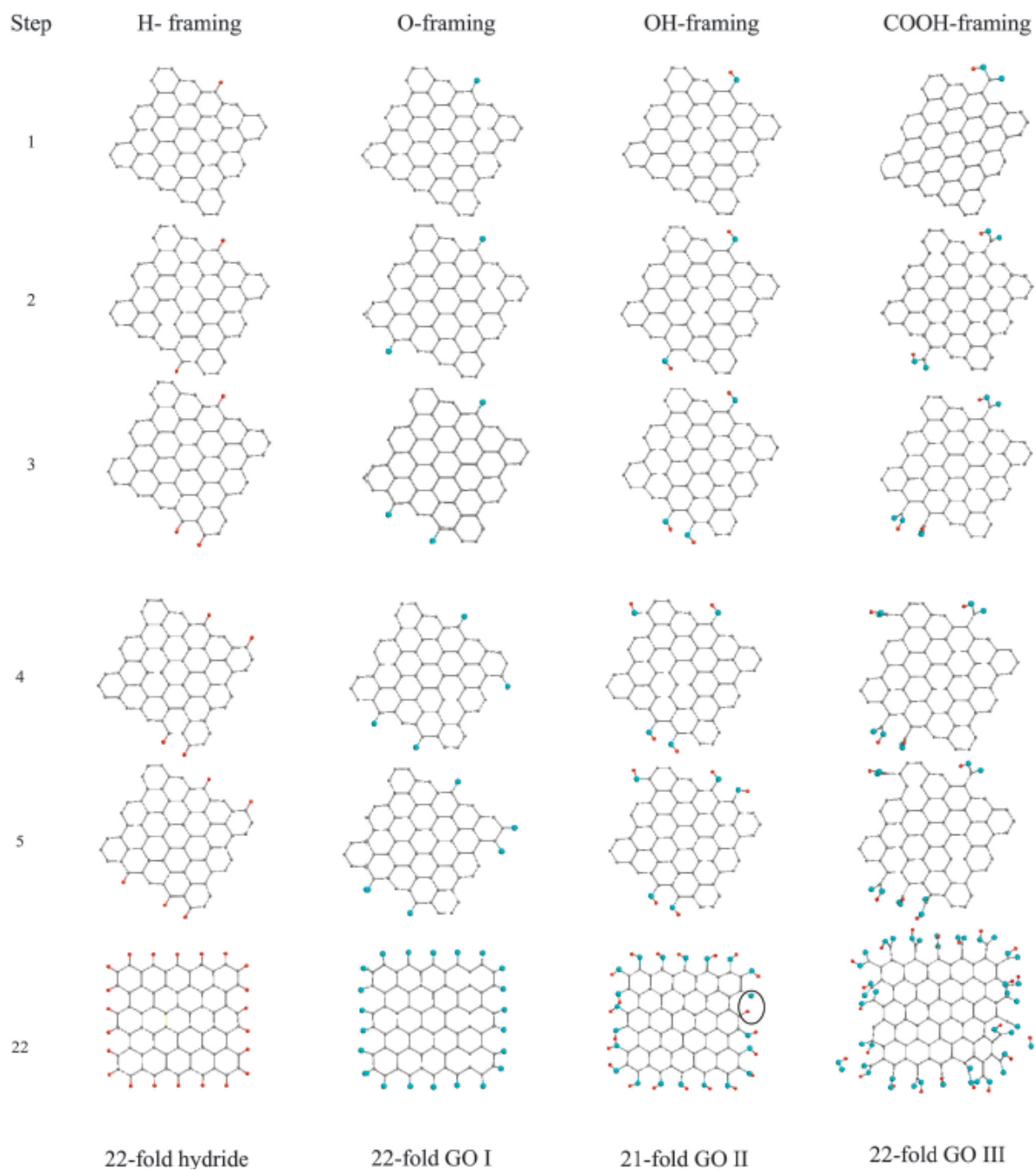

**Figure 8.** Equilibrium structures of (5,5) NGr polyhydrides and polyoxides related to the 1st, 2nd, 3rd, 4th, 5th, and 22nd (21st in the case of OH-framing, see text) steps of first-stage reactions. Roman figures mark the obtained GOs. Red and blue balls mark hydrogen and oxygen atoms, respectively [28].

Here, $\Delta H(GO_n)$, $\Delta H(GO_{n-1})$, and $\Delta H(Oxd)$ are heats of formation of the considered GOs at the $n^{th}$ and $(n-1)^{th}$ steps of oxidation, as well as lowest-total-energy (LTE) of oxidant itself, respectively.

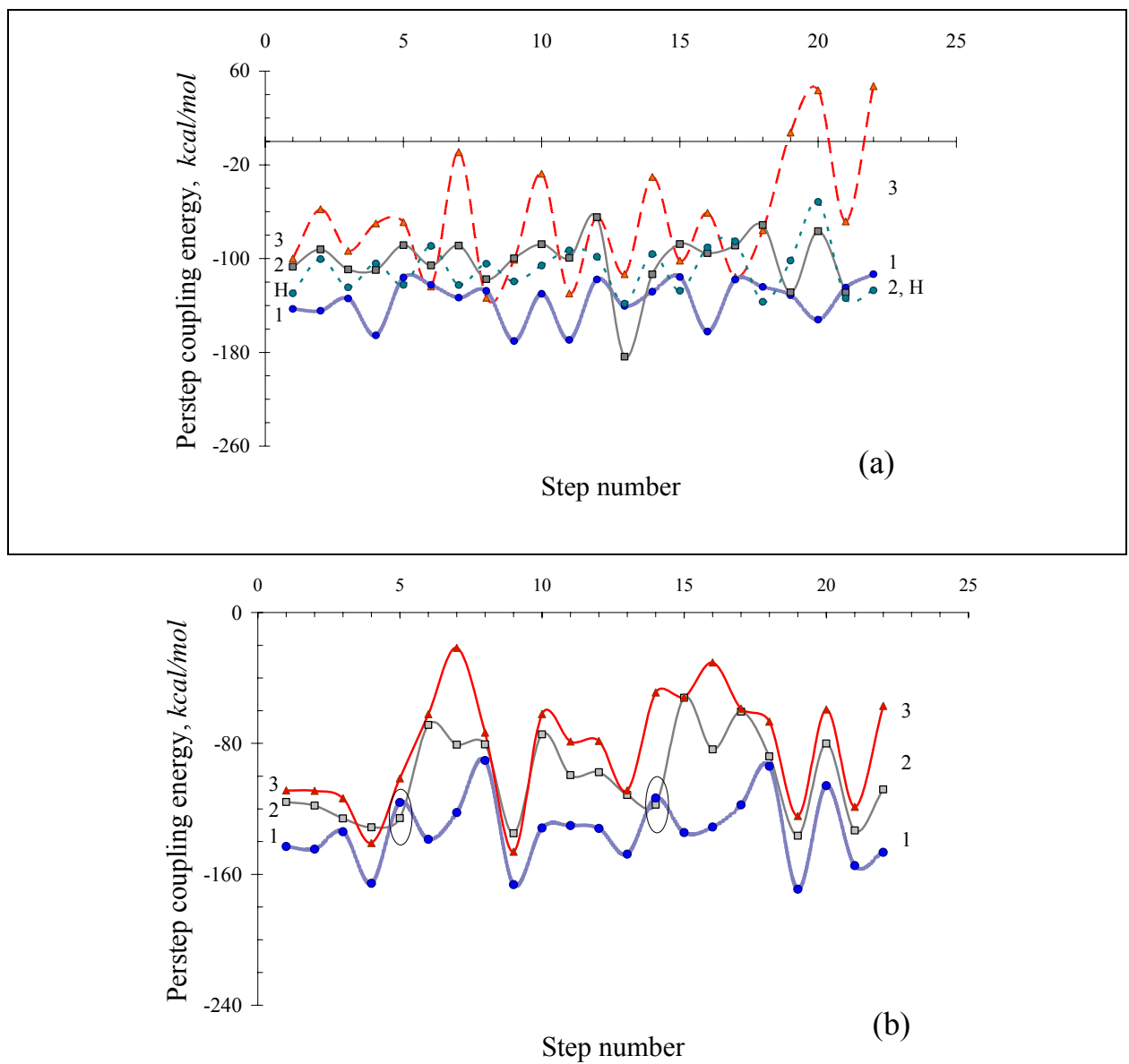

**Figure 9**. Per step coupling energy versus step number for families of GOs obtained in the course of the first- stage oxidation. (a). Homo-oxidant reactions, per step addition of O (curve 1, GO I), OH (curve 2, GO II), COOH (curve 3, GO III); H. Hydrides. (b). Hetero-oxidant reactions for GO V, per step addition of O (curve 1), OH (curve 2), and COOH (curve 3). Ovals mark intersections of curves 2 and 1 when OH-addition becomes more profitable than O-addition.

Analyzing data presented in Fig.9a, one can conclude the following.

1. The dependences of $E_{cpl}^{pst}(GO_n)$ on step number have much in common for all the oxidants. Thus, they show non-regular oscillating behavior that is characteristic for the hydrogenation as well once provided supposedly by a particular topology of the (5, 5) NGr molecule that reflects a changeable disturbance of the molecule carbon skeleton in the course of the chemical modification.

2. The molecule framing by carbonyls is evidently the most energetically favorable while oppositely, COOHs provide the least stable configurations. The coupling characteristic for OH-GOs is in between these two limit cases and only at the 13th step of the OH-GO is the largest one among the others due to dissociation of the hydroxyl group.

According to this finding, one should expect that carbonyls play the dominant role in the framing of GO molecules in practice. In contrast, the most spread opinion supports models of the GO chemical composition suggested by Lerf and Klinovsky [29, 30] where not carbonyls but carboxyls and epoxy groups frame the GO platelets. However, as shown in [27, 28], the epoxy-GOs are not only the least stable by energy but even become energetically non-profitable since PCE becomes positive from the third step.

### 4.1.2. Hetero-Oxidant Action

In practice, the homo-oxidant treatment of graphene has been so far observed for the graphene hydroxylation only [31]. The 21-fold GO II presented at the bottom of Fig. 8 is well consistent with the reported behavior of graphene oxide [31] and can be considered as a reliable model of the oxide structure. In all other cases, graphene oxidation predominantly occurs in the multi-oxidant media [3] so that the final products might present hetero-oxidant GOs. As suggested [27, 28], products synthesis can be traced in such a way. Basing of the ACS-governing algorithm, the first-attack target atom is determined and successively subjected to the addition of either O, or OH and COOH oxidant. GO which meets the requirement of the largest PCE criterion is selected and its ASC map is used to determine the second-step attack atom. Afterwards, the procedure of the successive addition of the three oxidants is repeated and new GO with the largest PCE is chosen. Its ACS map shows the target atom for the next attack and so forth. Figure 9b exhibits PCE plottings versus step number for the three oxidants. Comparing the data with those related to a homo-oxidant action shown in Fig. 9a one can conclude the following.

1. Plottings related to each of the considered oxidants are quite different in the two cases. This is a consequence of the difference of the chemical composition at the edge atoms during homo- and hetero-oxidant action at the current step.

2. Hetero-oxidant computational experiment exhibits quite undoubtedly that COOHs are not favorable for the (5, 5) NGr molecule framing.

3. The molecule framing is predominantly provided by carbonyls and partially by hydroxyls so that the final framing composition is double-oxidant corresponding to the 22-fold GO V (the oxide numeration follows [27, 28]) whose equilibrium structure is presented in Fig.10.

4. The carbonyl : hydroxyl ratio of 20:2, as well as the absence of COOH ending groups, are evidently non-axiomatic and quite strongly depend on the pristine graphene molecule size, as it will be shown in Section 5.1. However, dominance of carbonyls in the circumference area of GO molecules should be expected and seems realized in practice.

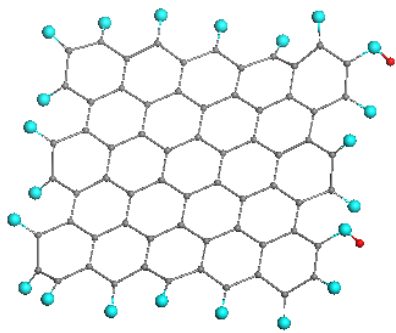

**Figure 10**. Equilibrium structure of free-standing (5,5) NGr membrane subjected to hetero-oxidant framing (equilibrium structure of GO V).

### *4.2. The Second-Stage Oxidation of the (5, 5) NGr Molecule*

The data presented in the previous Section, even once limited to the first steps of oxidation, convincingly show how complicated is the chemical modification of graphene in this case. Surely, the process does not stop at this level and continues further involving both edge atoms, if they are still not valence compensated, and basal ones thus becoming more and more complicated.

#### 4.2.1. Oxidation of Hydroxyl-Framed GOs

Figure 11 summarizes structural results typical for the second-stage homo-oxidant reaction related to a complete hydroxylation of the pristine (5, 5) NGr molecule. The 74-fold GO VII shown in the figure corresponds to the oxide equilibrium structure after termination of the 52nd step of the second-stage reaction. All additions occurred at basal plane and were considered for up and down hydroxyls positions. The GO best configuration was chosen according to the largest PCE criterion [27, 28]. Dense covering of the basal plane by hydroxyls has resulted in a considerable distortion of the carbon skeleton structure (Fig.11b) due to the reconstruction of flat benzenoid-packed configuration by corrugated cyclohehanoid-packed one. A large family of cyclohexane isomorphs may explain a fanciful character of the skeleton structure. In contrast to the (5, 5) NGr hydrogenation [22], it was unable to distinguish particular kinds of cyclohexane isomorphs like arm-chair and boat-like that were observed at the hydrogenation of two-side accessible free standing molecular membrane. None can exclude that the cyclohexane isomorph definition is not suitable for the chemically modified graphene that offers a limitless number of possible accommodation to minimize energy losses.

Evolution of PCE in due course of the 52-step reaction at the second stage is shown in Fig.11c. As seen, oscillating behavior observed for PCE during the first stage of oxidation is retained in this case, as well. When comparing the data with the PCE plotting related to GO II (curve 2), one can see a drastic difference in the absolute PCE values in the two cases. Actually, the average values for GO II and GO VII constitute -102.15 and -44.79 kcal/mol, which indicates two-and-half strengthening of hydroxyl coupling at their single addition to the molecule edge atoms in comparison with either the double additions to the edge atoms or single addition to atoms of the basal plane.

The plotting on the top of Fig.11c exhibits the evolution of the correlation of the (5, 5) NGr odd electrons in the course of the oxidation presented by changing the molecular chemical susceptibility $N_D$. The plotting shows that as the number of the attached hydroxyls increases, the number of effectively unpaired electrons, a total of 17.62 e for the pristine GO II, gradually decreases, thus highlighting gradual depletion of the molecular chemical ability. At the 52nd step $N_D$ becomes zero and keeps the value afterwards, which means stopping the reaction at this step.

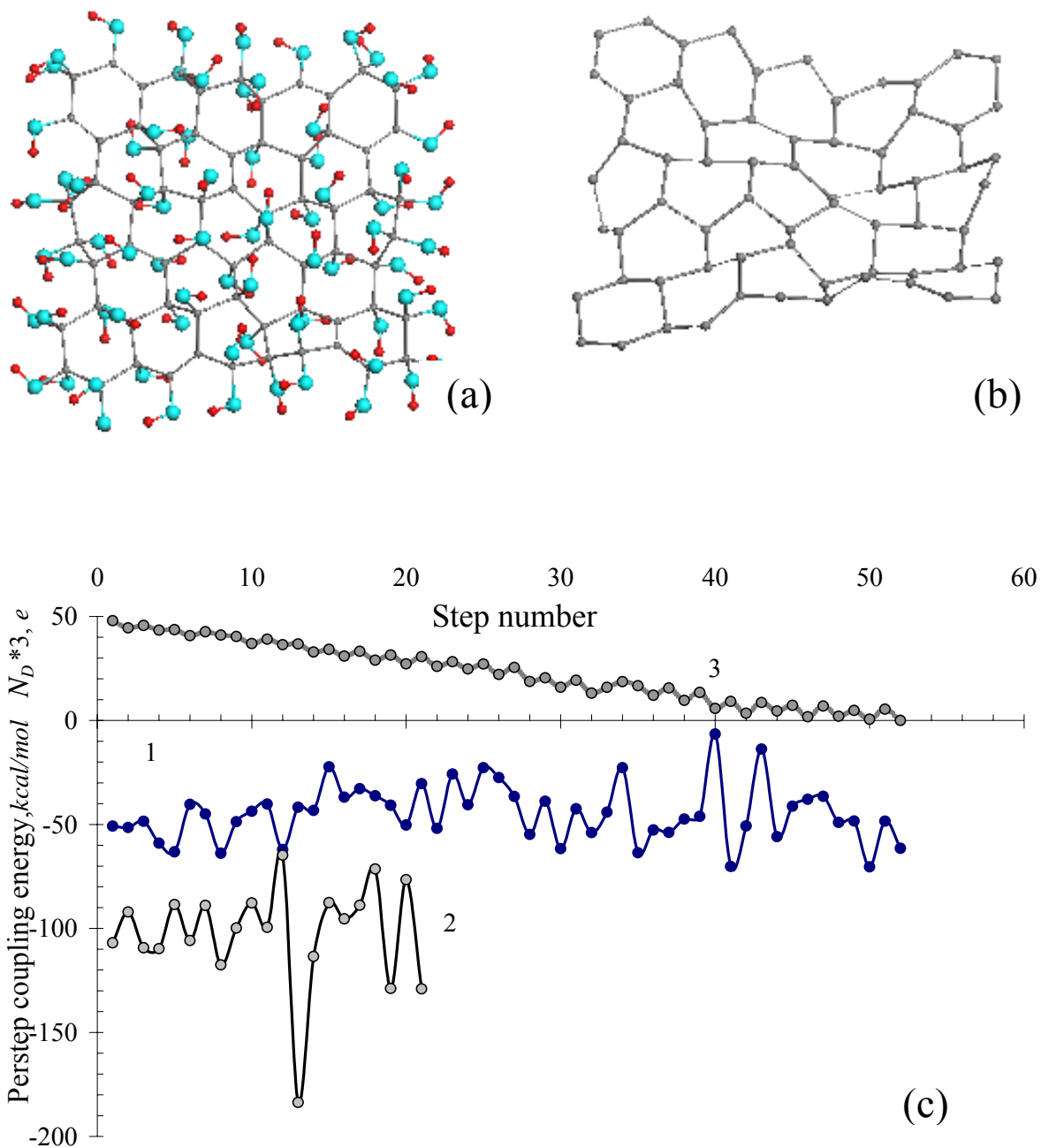

**Figure 11**. (a). Equilibrium structure of GO VII after 52 steps of the second-stage homo-oxidant reaction. (b). Core structure of GO VII. (c). Per step coupling energy versus step number for GO VII family in the course of the first- stage (1) and second-stage (2) oxidation. Top. Evolution of the total number of effectively unpaired electrons of GO VII during the second-stage oxidation.

4.2.2. Oxidation of Hetero-Oxidant-Framed GOs

Equilibrium structure of the final configuration of the 43-fold GO X obtained at the 21st step of the second-stage subsequent hetero-oxidant treatment is shown in Fig.12a. The final structure involves twenty carbonyls terminating the molecule edge atoms; four hydroxyls doing the same job, two of which are added during the second stage of the oxidation; fifteen epoxy and four hydroxyl groups randomly distributed over the basal plane. A particular attention should be given to a considerable curving of the carbon skeleton (Fig.12b) caused by the $sp^2$ - $sp^3$ transformation of the carbon valence electrons due to chemical saturation of their odd electrons. The structure transformation is similar to that one observed under one-side hydrogenation of (5, 5) NGr [22], albeit not so drastic. The considered configuration well suits real conditions, when one-side accessibility of the basal plane is the most favorable due to a top-down manner of graphite layer exfoliation [3, 32].

The PCE evolution, which accompanies the second-stage oxidation of GO V, is presented in Fig.12c. As seen in the figure, the O addition dominates and is characterized by an average PCE of -48,35 kcal/mol. The average PCE characteristic for hydroxyls constitutes -38,67 kcal/mol. For comparison, the average PCE for carbonyl-hydroxyl framing of GO V, presented by curve 3, is -133,76 kcal/mol.

*4.3. Main GO Characteristics and Properties Based on the Computational Experiment*

The discussed computational oxidation of the (5, 5) NGr molecule shows the following.

1. Two-zone chemical reactivity of the graphene molecule causes the two-stage character of the oxidation. The first reaction zone covers circumference area that includes edge atoms with high ACS values while the second zone involves all basal atoms as well as those edge atoms with much lower ACS whose chemical reactivity was not suppressed in the course of the first-stage reactions.

2. The first-stage reaction concerns 22-step framing of the (5, 5) NGr molecule by oxidants among which addressed oxygen atoms, hydroxyls and carboxyls. The obtained fully framed GOs involve both homo-oxidant and hetero-oxidant GOs. All the GOs correspond to one-point contact of oxidants with the molecule edge atoms thus providing the formation of C=O, C-OH, and C-COOH ending groups. Table 2 summarizes average PES and oxidant-induced chemical bonds data that characterize the obtained GOs.

3. The energy needed for each step of the second-stage oxidation decreases about three times for all the oxidants indicating that the oxidant coupling with the basal atoms is much weaker in comparison with the interaction occurred at edge atoms. The latter is supported by the elongation of all the second-stage formed C-OH bonds and the formation of long O-C bonds of epoxy groups in comparison with rather short C=O bonds of ending carbonyls.

4. The obtained results allowed suggesting a reliable vision of the GO composition that meets the requirement of the top-down oxidation and is energetically profitable: GO X presented in Fig.12a seems to be one of most reliable model structures for GOs obtained by one of the versions of the Hummers method [3]. Computations have shown that carbo-carboxyl contacts are the least probable at both the first and the second stages of oxidation so that if even present they constitute a definite minority in the GO final structures.

GO X

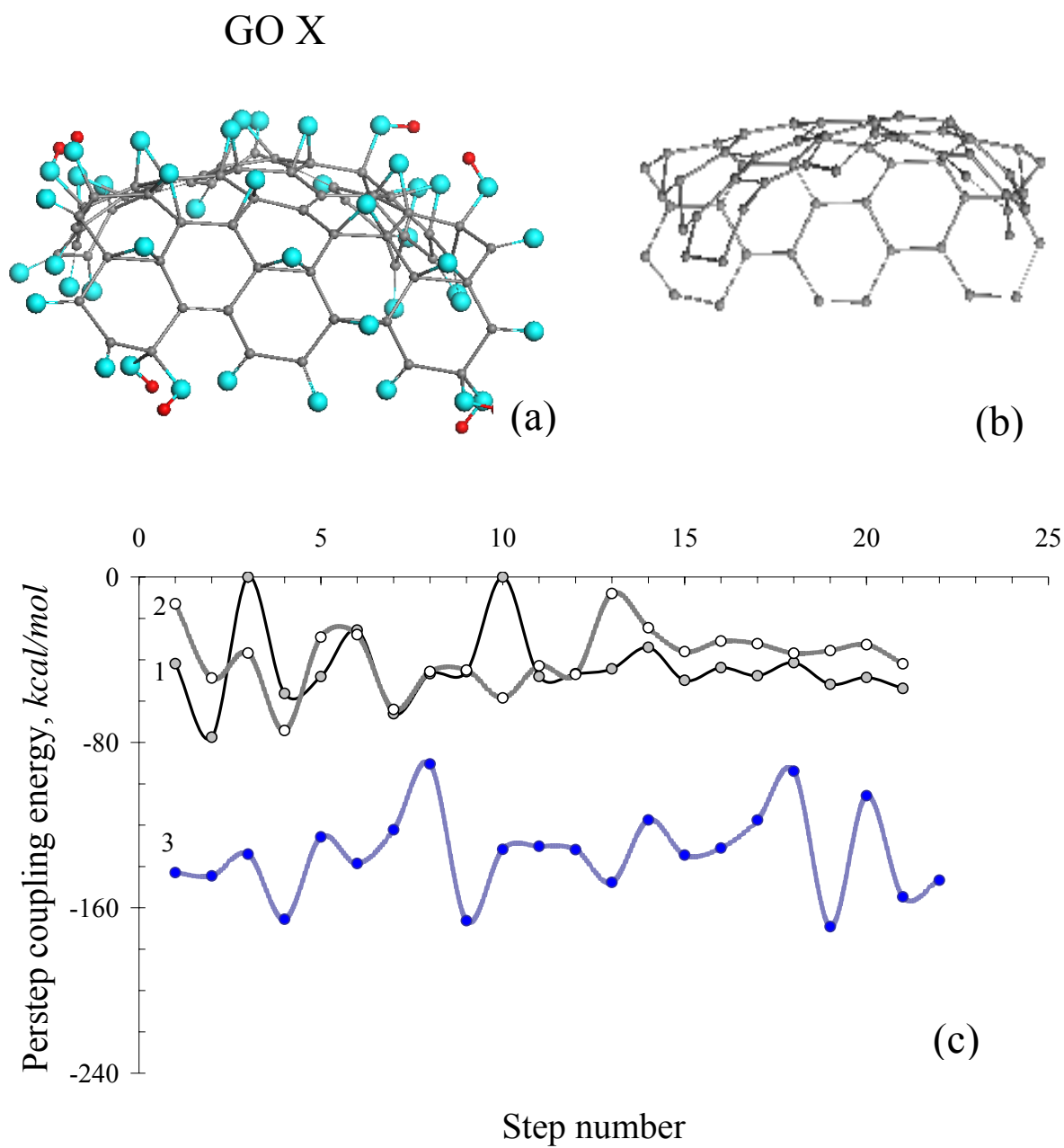

**Figure 12**. (a). Equilibrium structure of GO X obtained in the course of hetero-oxidant second-stage oxidations. (b). Core structure of GO X. (c). Per step coupling energy versus step number for GO X family under O (curve 1)- and OH (curve 2) -additions in the course of the second-stage oxidation. Curve 3 plots the same data for GO V family during the first-stage hetero-oxidant reaction.

5. Concerning GO morphology, it should be noted that the performed calculations have not revealed either regularly structured GOs or even those close to regularly structured ones.

**Table 2**. Average characteristics of the (5, 5) NGr molecule oxidation [27]

| GOs[1] | PCE, *kcal/mol* | | | C-O bonds lengths, *Å* | | | C-H bonds, *Å* |
|---|---|---|---|---|---|---|---|
| | O | OH | COOH | C-OH | O-C-O | C=O | |
| The first-stage oxidation | | | | | | | |
| I | -135,65 | - | - | - | - | 1,234 | - |
| II | - | -102,15[2] | - | 1,372 | - | 1,249[3] | 1,101[3]; 0,971 |
| III | - | - | -77,34 | 1,361 | - | 1,233 | 0,972 |
| V | -133.14 | -101,08 | -82,77 | 1,348 | - | 1,233 | 0,978 |
| The second-stage oxidation[4] | | | | | | | |
| VII (IV)[5] | - | -44.79 | - | 1,371 1,417 | - | - | 0,969 |
| X (V) | -48,35 | -38,67 | - | 1,413 | 1,436 | 1,223 | 0,971 |

[1] Oxides under the corresponding numbers are presented in Figs. 8, 10, and 12

[2] The value is slightly overestimated due to one hydroxyl dissociation in due course of the reaction.

[3] Solitary C=O and C-H bonds, see Fig.8

[4] The PCE values are averaged over steps of the second stage only.

[5] Here and below figures in brackets point the reference to GOs of the first-stage reaction.

Moreover, the GO carbon skeleton has lost its planarity at the early stage of oxidation and becomes remarkably curved when the oxidation is terminated. Both features are due to $sp^2$- $sp^3$ transformation of the carbon electron system and the substitution of flat benzenoids of graphene by corrugated cyclohexanoids of GOs. A large variety of cyclohexanoid structures greatly complicates the formation of regularly structured graphene derivatives, in general, which is why the regular structure of graphene polyderivatives should be considered as a very rare event. Until now, it has been observed so far for a particularly arranged graphene hydride named graphane [22].

### *4.4. Oxidation-Induced Structure Transformation*

The stepwise oxidation is followed by the gradual substitution of the $sp^2$-configured carbon atoms by $sp^3$ ones. Since both valence angles between the corresponding C-C bonds and the bond lengths are noticeably different in the two cases, the structure of the carbon skeleton of the pristine (5,5) NGr molecule loses its flatness and becomes pronouncedly distorted. Comparing the pristine bond length diagram with those belonging to a current oxidized species makes it possible to trace changes of the molecule skeleton structure.

Figures 13a and b demonstrate the transformation of the skeleton structure in the course of both first-stage and second-stage oxidation exemplified by the changes within a fixed set of the C-C bonds of homo-oxidant hydroxylated GOs II and VII. As seen in Fig.13a, the OH-framing of GO II causes a definite regularization of the C-C bonds distribution. None of single C-C bonds is formed since the chemical activity of edge atoms is not completely suppressed due to which $sp^3$-$sp^2$ transformation does not occur. Contrary to this, the second-stage oxidation is followed by the formation of single C-C bonds that evidently prevail in Fig.13b thus exhibiting a large-scale $sp^3$-$sp^2$ transformation. The appearance of the elongated C-C bonds, the number of which increases when oxidation proceeds, is naturally expected. However, to keep the skeleton structure integrity, this effect as well as changes in valence angles should be compensated. At the level of bonds, this compensation causes squeezing a part of pristine bonds. The plotting in Fig.13b shows seven C-C bonds which are extremely short and whose lengths are in the interval of 1.345-1.355 Å. These lengths point to non-saturated valence state of the relevant carbon atoms which are not attached by oxygen, on one hand, however, on the other hand, are under the critical $R_{crit}$ value of 1.395Å that put a lower limit for observation of the odd electron correlation. Under this value, the bond length provides a complete covalent bonding of odd electrons due to which these electrons are non-correlated so that the total number of effectively unpaired electrons $N_D$ is zero [2]. The formation of these short bonds explains stopping of the oxidation reaction at the 52[th] step of the second-stage reaction.

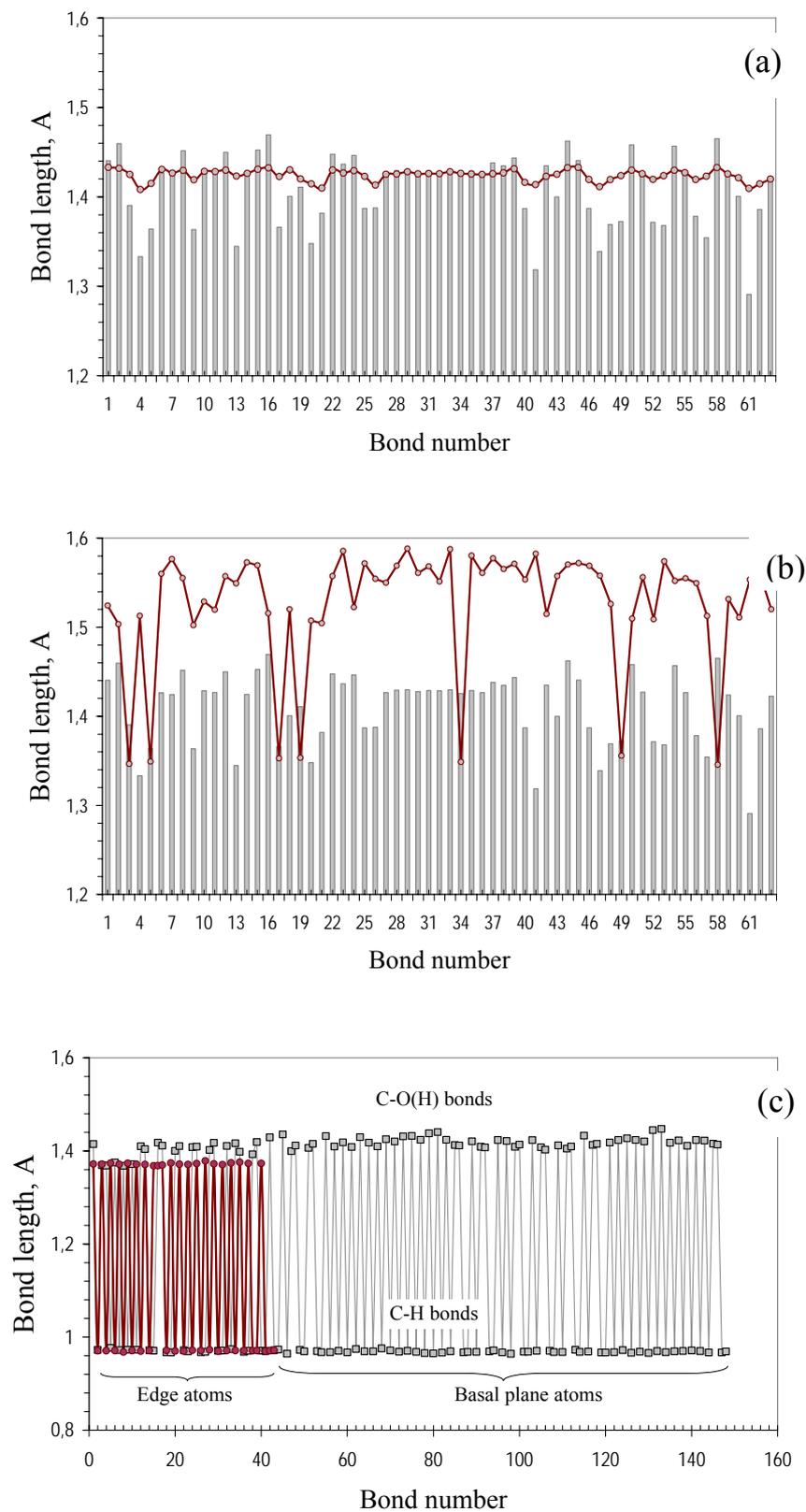

**Figure 13**. The $sp^2$-$sp^3$ transformation of the (5, 5) NGr molecule skeleton structure in the course of the successive homo-oxidant oxidation; the C-C bond length distribution after the first-stage oxidation, GO II (a) and the second-stage oxidation, GO VII (b). Light gray histogram is related to the pristine molecule. c. Oxidant-given C-O(H) and C-H bonds at the first-stage (dark red) and the second-stage (gray) oxidation.

Figure 13c presents the bond length distribution related C-OH and O-H oxidant groups. Dark red plotting is related to GO II and describes the bonds of framing atoms while gray plotting shows similar distribution in the case of the 74-fold GO VII. As seen in the figure, the bonds are quite regularly distributed in the both cases. The relevant dispersion of the bonds is listed in Table 3. As seen in the figure, all the C-OH bonds related to hydroxyl attached to the basal plane atoms are longer than those related to OH-framing of the molecule during the first stage of oxidation (1.417 and 1.371Å, respectively). Important to note that previously short C-OH bonds related to edge atoms elongate during the second stage of oxidation as well and approach the average length of 1.417 Å. At the same time, part of these bonds, which remain untouched during this stage, preserve their shorter lengths.

The transformation of the molecule skeleton structure in the case of the hetero-oxidant reaction is shown in Fig.14. Figure 14a presents the skeleton distortion of GO V by the end of the first-stage oxidation. Since 20 carbonyls and two C-OH groups form the molecule framing, the dominant elongation of C-C bonds due to $sp^3$-$sp^2$ transformation provided by carbonyls is clearly seen Two C-OH groups leave the corresponding edge atoms in the $sp^2$ configuration. The second-stage oxidation concerns two edge and 38 basal atoms. Shown in Fig.14b exhibits the skeleton distortion by the 21[st] step of the second-stage oxidation for GO X. The dominant majority of bonds become single due to the large-scale $sp^3$-$sp^2$ transformation. The remaining untouched carbon atoms form bonds of 1.343-1.358Å in length similarly to the case shown in Fig.13b. Reasons for these bonds creation are discussed above. As previously, the bonds length provides a complete covalent bonding of the remaining odd electrons thus conserving $sp^2$ configuration of the relevant carbon atoms and terminating further oxidation. The appearance of shortened C-C bonds in GO VII and GO X, lengths of which are less than the critical value of 1.395Å, is quite similar to that takes place during the stepwise hydrogenation [25] and fluorination [21] of fullerene $C_{60}$.

Figure 14c discloses events that concern attached oxidants. As said above, the first-stage oxidation is terminated by a complete framing of the pristine molecule by 20 carbonyls and two hydroxyls. Consequently, dark red curve in Fig.14c exhibits 20 C=O bonds as well as two C-OH and two O-H bonds of GO V. Gray curve presents the bond structure after terminating of the second-stage oxidation for GO X at the 21$^{st}$ step. As seen, previously much shorter C-OH bonds of 1.348Å lengthen up to 1.413Å and are comparable by length with all the newly formed C-OH bonds; C=O bonds differ slightly, and C-O-C bonds of epoxide groups are added. The average length values are listed in Table 3 alongside with the bond length dispersion. As seen in the table, the bond distributions are quite homogeneous, and their dispersion does not exceed 1%.

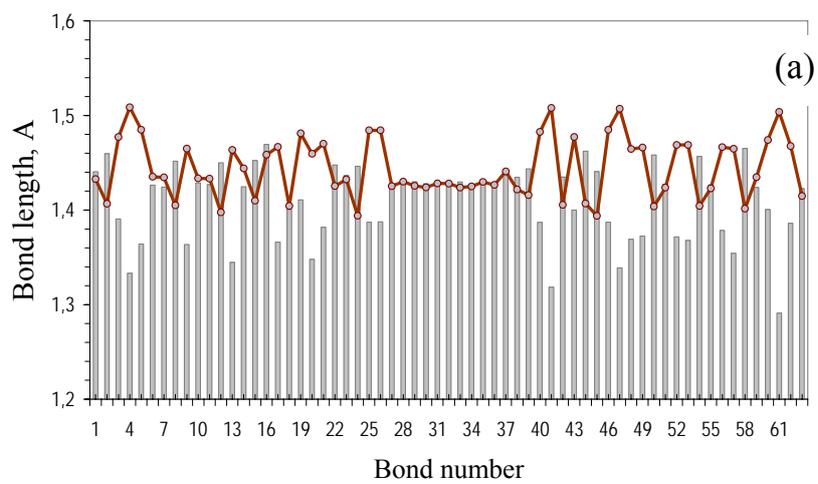

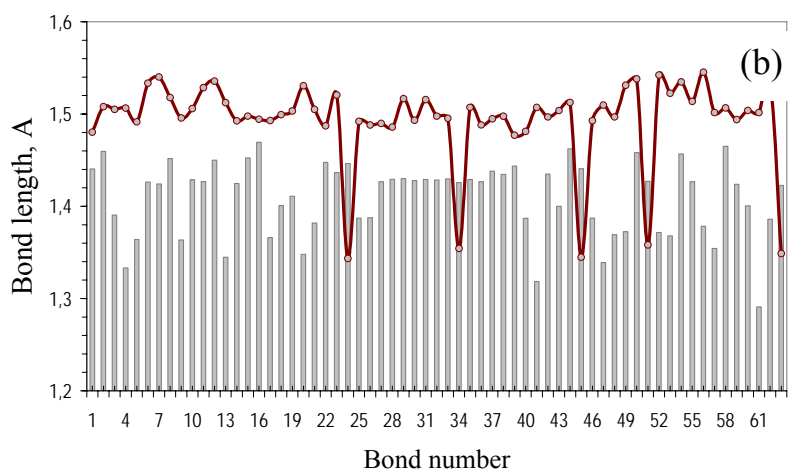

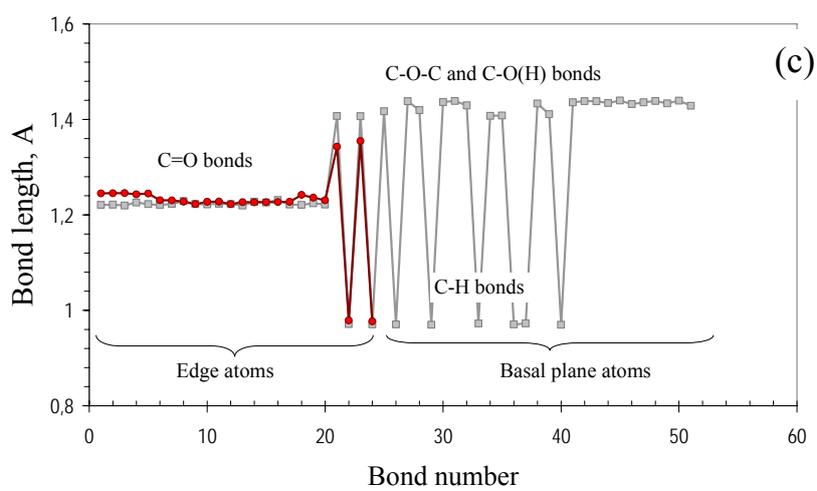

**Figure 14.** The $sp^2$-$sp^3$ transformation of the (5, 5) NGr molecule skeleton structure in the course of the successive hetero-oxidant oxidation; the C-C bond length distribution after the first-stage oxidation, GO V (a) and the second-stage oxidation, GO X (b). Light gray histogram is related to the pristine molecule. c. Oxidant-given C=O, C-O-C, C-OH and O-H bonds at the first-stage (red) and the second-stage (gray) oxidation.

**Table 3**. Average bond lengths of the (5, 5) NGr oxides and their dispersion, % [27]

| GOs[1] | C-O bonds lengths, $Å$ | | |
|---|---|---|---|
| | C-OH | O-C-O | C=O |
| First-stage oxidation | | | |
| I | - | - | 1,234 (5.30; - 0,90) |
| II | 1,372 (0.70; -0,30) | - | 1,249[2] |
| III | 1,362 (3.79; -0.99) | - | 1,233 (0.58; -0.47) |
| V | 1,348 (0.43; -0.43) | - | 1,233 (1.04; -0.83) |
| Second-stage oxidation | | | |
| VII (IV)[3] | 1,371 (0.32; 0.18) 1,417 (1.98; -1.68) | - | - |
| X (V) | 1,413 (1.08; -0.46) | 1,436 (0.24; -0.46) | 1,223 (0.63;-0.30) |

[1] Oxides under the corresponding numbers are presented in Figs. 8, 10, 11a, and 12b.
[2] A solitary C=O bond, see Fig.9
[3] Here and below the figures in brackets point the reference to GOs of the first-stage reaction.

### *4.5. Some Comments Concerning GO Reduction*

The reduction of massively produced GOs is the second chemical reaction on the way of the transformation of dispersed graphite into a powder consisting, in an ideal case, of one-layer graphene flakes. Looking at GO VII (Fig.11a) and GO X (Fig. 12a) as reliable models of GO in the case of homo- and hetero-oxidant processes and aiming at their returning to the graphene molecule shown in Fig.2a, one should answer the following questions:

    1. Is it possible to return a drastically deformed and curved carbon skeleton structures of GO VII and GO X to the planar structure of the (5, 5) NGr molecule?

    2. Is it possible to take out all the oxidant atoms from GO VII and GO X making them free of dopants?

       Core structures X and $X_{fr}$ shown in Fig.15 correspond to GO X subjected to optimization after removing all the oxidant atoms, in the first case, and oxidant atoms attached to the molecule during the second-stage of oxidation only, in the second. Skeleton structures VII and $VII_{fr}$ in the figure demonstrate the GO VII cores after similar removing concerning OH groups. As seen in the figure, the optimization/reduction of cores X (Fig.12b) and VII (Fig.11b) fully restores the flat structure of the (5, 5) NGr molecule. The difference in the total energy of thus restored structures constitutes 0, 3% and 1.1% the energy of the (5, 5) NGr molecule relating to cores X and VII. Therefore, a drastic deformation of the GOs carbon skeletons is no obstacle for the restoration of the initial flat graphene pattern in spite of large deformational energy, namely, 874 kcal/mol and 1579 kcal/mol, incorporated in the deformed X and VII core structures. This is an exhibition of the extreme flexibility of the graphene structure that results in a sharp response by structure deformation on any external action, on one hand, and provides a complete restoration of the initial structure, on the other.

    It should be noted, however, that not pure graphene but slightly oxidized product, called as rGO (see [3, 32-36] and references therein), is obtained after the GO reduction in practice. As follows from the plottings in Fig.13c, GOs are characterized by two regions of chemical bonding of oxidants with graphene body. While the edge atoms region should be attributed to that one of strong chemical bonding, the basal plane, for which the average coupling energy is three times less, is evidently related to the area of a weak coupling. The finding is crucial for the GO reduction showing that the latter concerns basal plane in the first instance while oxidants located in the framing area may not be removed under conditions of the convenient reduction without destruction of the carbon skeleton. This finding explains the residual oxygen content in reduced rGOs of 5-10% [32, 37-39]. The equilibrium structure of core $X_{FR}$ in Fig. 15 may give a reliable presentation of the product while core $VII_{fr}$ may be attributed to the main product obtaining during reduction of the hydroxylated graphene [31].

### 5. How Does Chemical Modification of Graphene Molecules Affect Electronic Structure?

Physical impact of the graphene chemical modification has been largely discussed (see 4, 40 and references therein). The main issues concern opening the gap in the energy spectrum of graphene crystal and peculiar edge atom states in the vicinity of the Fermi energy. The first is caused by the violation of the crystal planarity due to the $sp^2$-$sp^3$ transformation of carbon atoms. The

GO core

X

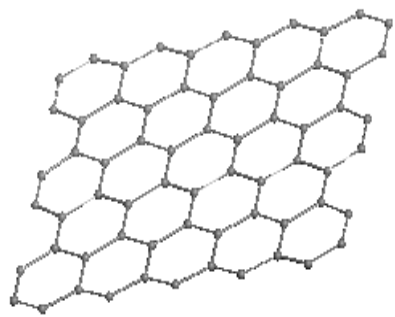

X$_{fr}$

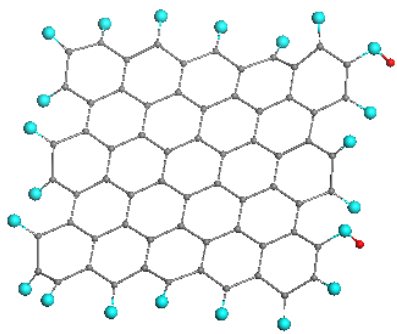

VII

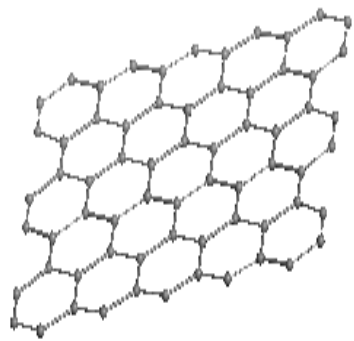

VII$_{fr}$

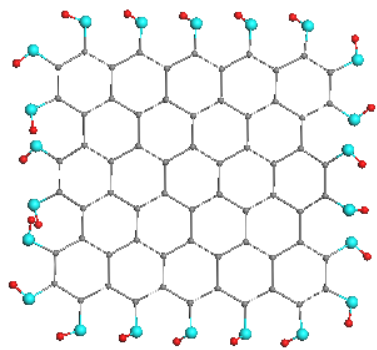

**Figure 15**. Core structures of GO X  (X and X$_{fr}$) and GO VII ( VII and VII$_{fr}$) after (b) structure optimization (see text).

second, according to R. Hoffmann [41], are 'the origin of the reactivity of the terminated but not passivated point, line, or plane' but 'may be mistaken for real states in the band gap, important electronically'. In spite of physical consequences, the chemical nature of both effects is clearly vivid.

Serious consequences caused by chemical modification are not limited to peculiar physical properties of graphene crystal but play a significant role in the chemical behavior of the substance. The latter concerns the size-dependent chemical composition of final products of chemical reactions, destruction of graphene samples in the course of chemical reactions, changing the character of mechanical behavior of graphene as well as of other topochemical reactions, and much more. Below a few examples will be considered.

### 5.1. Size Dependence of Chemical Reactions Final Products

As shown in the preceding Sections, the circumference of any graphene molecule governs any chemical reaction determining the reaction flowing from the very beginning. At the same time, the chemical reactivity of the area changes in the course of reaction due to a highly sensitive response of the molecule carbon skeleton structure to each act of chemical addition. In its turn, the structure reconstruction results in a redistribution of C-C bonds over length thus promoting redistribution of the atomic chemical susceptibility over the molecule atoms. Due to high restructure ability, both progress of reaction and final chemical composition of its product obviously depends on the molecule size. Intensively studied chemistry of graphene oxides is strong evidence of the correctness of this statement. Enough to mention ongoing debates about the chemical composition of the GO circumference area, concerning the presence of carboxyl units, in particular. One of the reasons for the question not to be solved until now is that experimental reports on FTIR, XPS, Raman scattering, optical spectra, and so forth presented by different groups of chemists are not identical in spite of using similar technological procedures for the GOs production. The ambiguity may be understood if suppose that the final product in each case presents dispersion of GOs differing by size and chemical structure. In its turn, size-distribution profiles of the dispersions are obviously sensitive to the reaction conditions that might be different. It is obvious that what is said can be attributed to the reduced graphene as well. As for the latter, the optical spectroscopy of graphene quantum dots, which present dispersions of rGOs, gives a straight evidence of the structural and electronic inhomogeneity of the dots (see review [42] and references therein).

Figure 16 presents results of quantum chemical calculations performed for the current chapter. The equilibrium structure of the (5, 5) rGO molecule shown in panel *a* has been discussed in Section 4.5. The structure of the (11, 11) rGO molecule in panel *b* was obtained in the course of stepwise hetero-oxidant addition of three oxidants to the pristine (11, 11) NGr molecule (see Fig. 17a) that is twice and four times larger the former one by linear size and atom number, respectively. At each reaction step, the choice in favor of oxidant, which contributes to the final product, was made on the basis of the largest PCE, as was considered in Section 4 in details.

As seen in the figure, the chemical composition of the both molecules framing is quite different. Twenty carbonyls and two hydroxyls compose the framing in the first case while thirty carbonyls, nine hydroxyls and 5 carboxyls frame the second molecule. The difference is well explained by PCE behavior in both cases (see Fig. 9b and 16c). In all the cases, the PCE plottings of both molecules for all oxidants oscillate around a mean value. In spite of that the average PCE values differ considerably for the oxidants, large amplitudes of the plottings cause their intersection due to which carbonyls are substituted by not only hydroxyls as in the case of the (5, 5) rGO molecule but by carboxyls as well in the case of (11, 11) rGO. This substitution can be traced when following along the red curve in Fig. 16c. Therefore, the oxidant composition of the circumference area of both GO and rGO molecules depends on the molecule size. However, there are serious reasons to expect the carbonyl contribution to be the largest by both absolute and relative values while the carboxyl one to be relatively much smaller and absolutely the least.

### 5.2. Chemically Induced Destruction of Graphene Samples

The chemistry of graphene oxidation has revealed one more size dependent effect. As follows from reviews [3, 30, 35, 37], the size of chemically produced GO is always less than that of the pristine graphite. Moreover, the GO size reduces when O:C at% ratio increases. As shown, oxidation causes destruction of the pristine graphite and graphene sheets just cutting them into small pieces [34, 38]. Thus, 900 sec of continuous oxidation cut a large graphene sheet into pieces of ~1 *nm* in size [38]. Important, that further prolongation of the oxidation does not cause decreasing the size thus stabilizing them at the 1 *nm* level. This finding has allowed the author to put forward a revolutionary idea about the real structure and origin of the well known natural allotrope of carbon – shungite – and present it as loosely packed fractal nets of graphene-based (rGO) quantum dots [43].

The feature is evidently connected with the role of the edge zone for the chemical interaction. If there is any topological reason for distinguishing a piece of the carbon atomic structure as an edge zone, strong chemical interaction with addend immediately fixes the area thus creating conditions for the zone enlarging due to anisotropy of the reconstructed odd electron cloud, on one hand, and mechanical stress, on the other.

A high ability of the graphene molecule to follow this way is distinctly revealed via its ACS map. It should be remained that changing in the ACS maps image reflects the changes in the C-C bond length distribution caused by the relevant chemical modification. Figure 17 presents a comparable view of the map image changing caused by one-atom hydrogenation of edge atoms of (5, 5) NGr, (11, 11) NGr, and (15, 12) NGr molecules. Carbon core of the latter is more than three and six times bigger the former one by linear size and atom number, respectively. As seen in the figure, the hydrogen addition created a vivid, highly anisotropic topological changing of all the molecules. However, if in the case of the first molecule, these changes do not seem to destroy the molecule, in the case of the other two molecules, new topological channels of drastically reduced chemical activity for some atoms and significantly enhanced on the other, extending through the middle of the molecules, undoubtedly indicate a possible division of the latter into two parts due to their chemical modification.

Apparently, this feature of the ACS maps lays the foundation of graphene molecules fragmentation during chemical reactions. The channels are characteristic for large molecule while small molecules can retain their integrity. Unfortunately, the picture typical for hydrogen terminating cannot be expanded over other chemical additions due to extremely strong dependence of the ACS map image on the chemical addend. This is well seen in Fig. 8 showing how significantly chemical addend influences the target-atom ordering in the course of stepwise reactions. Similar effects are seen in Figs.4 and 5 exhibiting a considerable reconstruction of the C-C bond distribution at each addition of one hydrogen atom. The odd electrons correlation, which provides collective behavior of the latter, forms the grounds for the feature. The latter makes difficult a prediction of the molecule fragmentation at different chemical reactions.

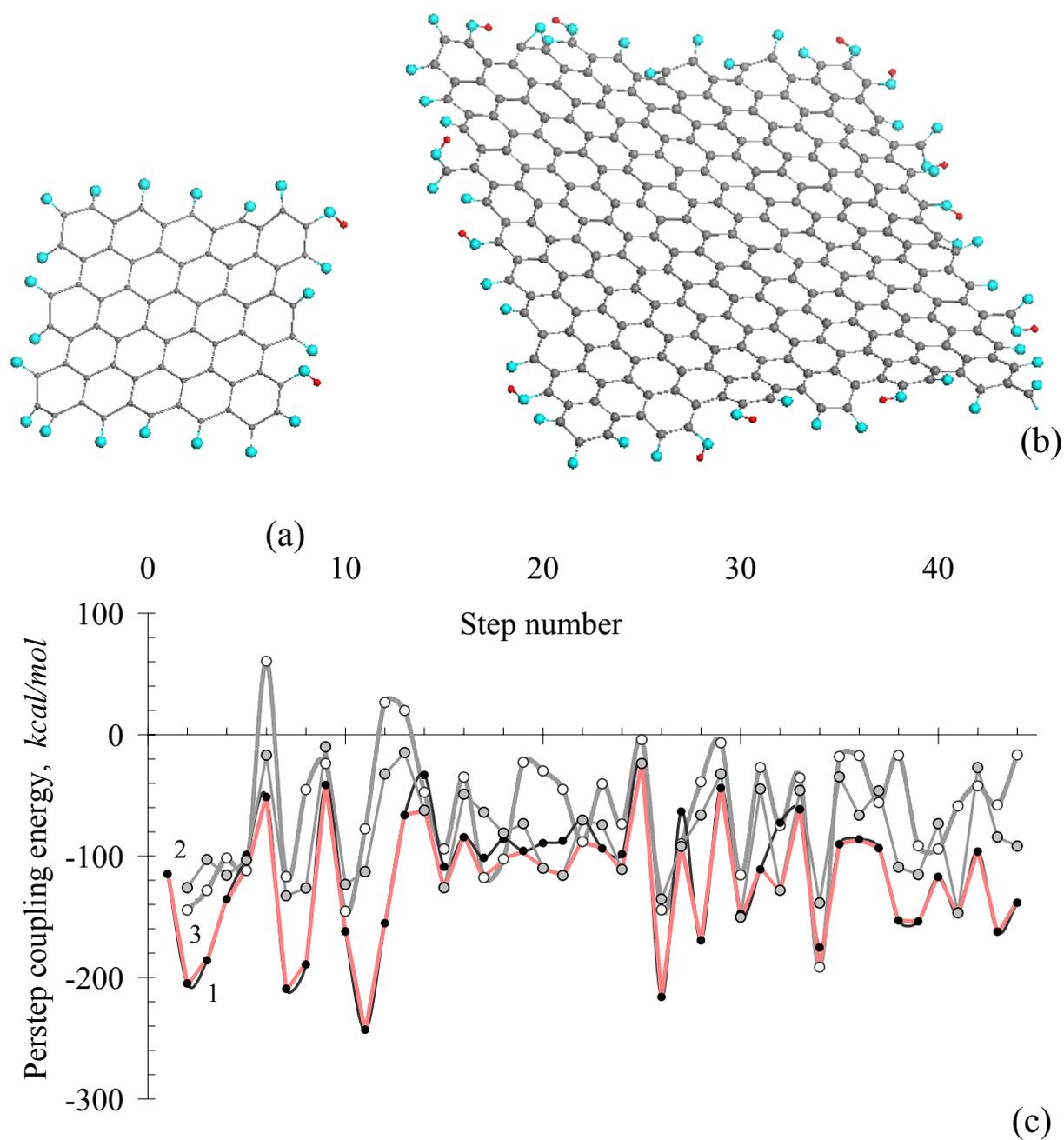

(a)

(b)

(c)

**Figure 16**. Equilibrium structure of (5, 5) rGO (a) and (11, 11) rGO (b) molecules obtained in the course of hetero-oxidant first-stage oxidations. (c). Per step coupling energy versus step number for GO XII family under O (curve 1)-, OH (curve 2)-, and COOH (curve 3)-additions. Red curve presents the PCE plotting that determines the formation of the (11, 11) rGO final structure.

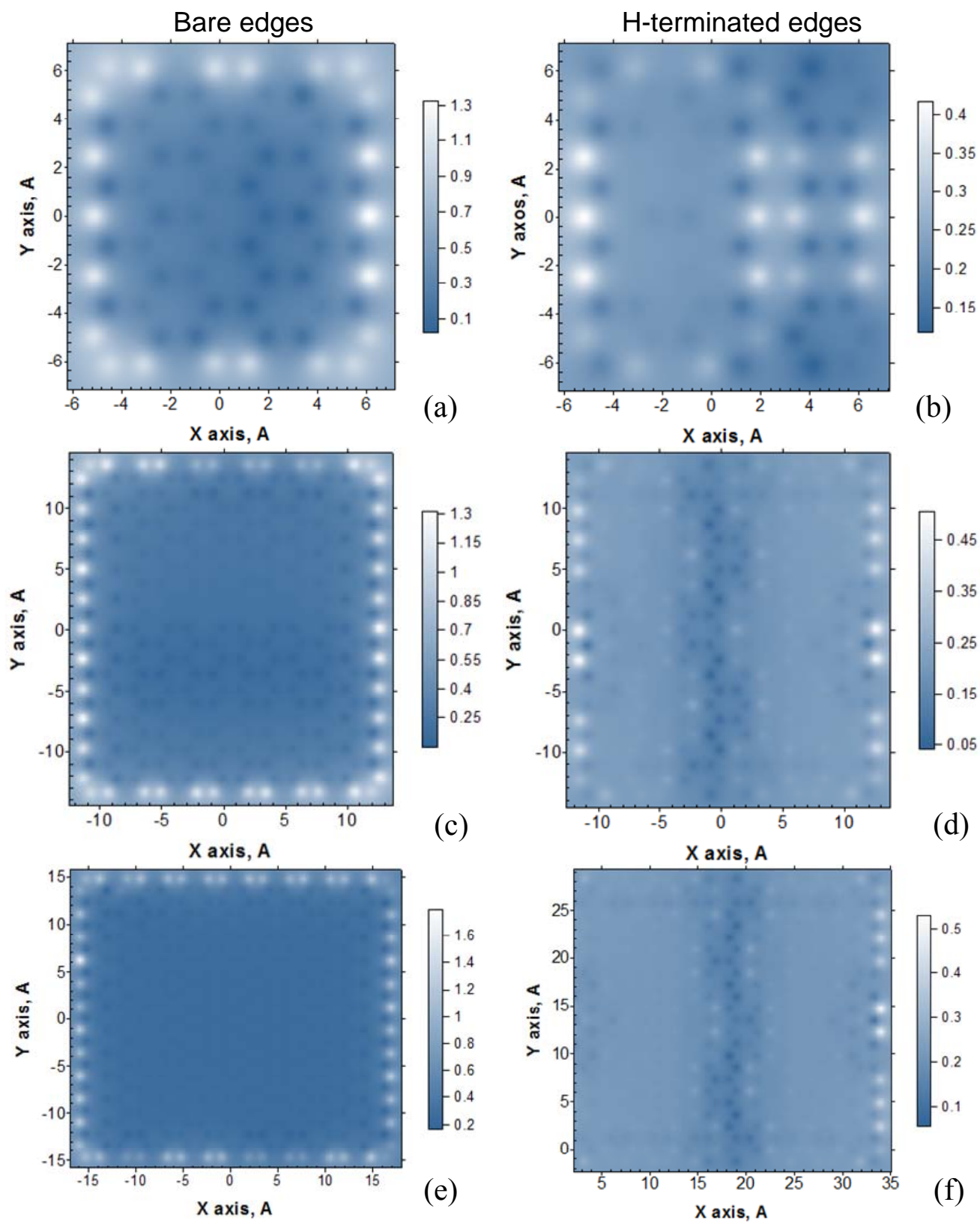

**Figure 17**. ACS image maps of the bare (left column) and terminated by hydrogen atoms per each edge atom (right column) of (5, 5) NGr (a and b), (11, 11) NGr (c and d), and (15, 12) NGr (e and f) molecules.

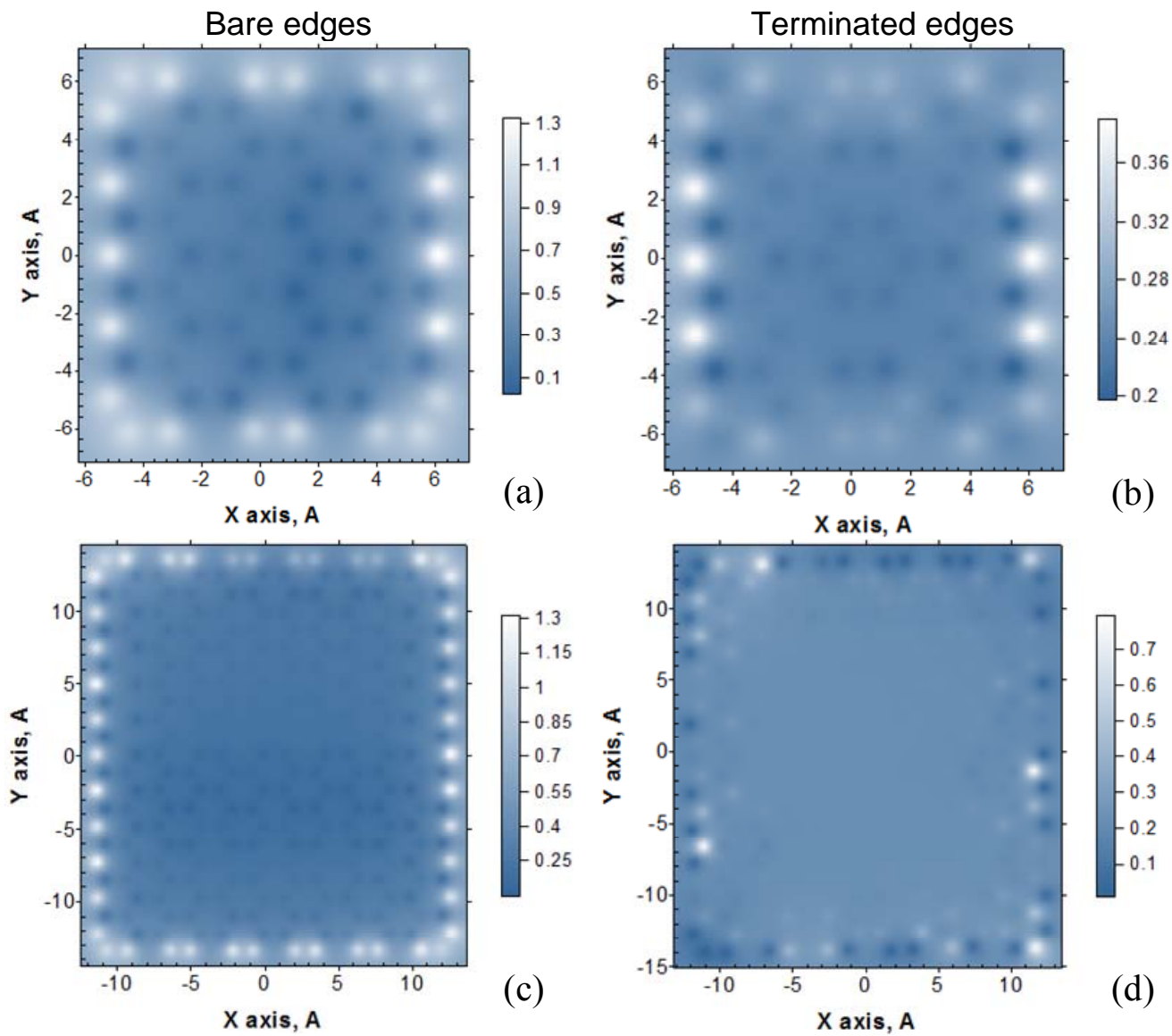

**Figure 18** ACS image maps of the bare (left column) and framed (right column) of (5, 5) NGr (a and b) and (11, 11) NGr (c and d) molecules. The framed molecules are presented by (5, 5) rGO (see VII$_{fr}$ in Fig. 15) and (11, 11) rGO in Fig. 16b.

Thus, Fig. 18 presents the ACS image maps of the (5, 5) NGr and (11, 11) NGr molecules subjected to complete first-stage oxidation. As seen from Fig.18b, the molecule edge hydroxylation that, apparently, was to be similar to monatomic hydrogen termination shown in Fig.17b, leads to absolutely different ACS map not showing any indication on topological changes over the molecule carbon skeleton. No similar indications follow from the ACS map of the (11, 11) rGO molecule that corresponds to the (11, 11) NGr molecule subjected to the hetero-oxidant first-stage oxidation. Since all the studies show that the odd electrons form an easily changeable, live system and since the mentioned topological channels are obviously created in the course of a collective process, it is difficult to predict at which namely step of reaction, say, oxidation, the disruption of graphene molecule will start. It is obvious that computationally the problem solution is greatly time consuming now. Some new algorithms might be needed to perform the job.

### 5.3. Chemically Affected  Topological Mechanochemistry of Graphene

Graphene molecule is a very interesting topological object [44] and its chemical modification allows for making the feature more bright [45].  Below we will consider a particular topological effect caused by the influence of both the loading direction and the graphene molecule edge termination on the inherited topology of the molecule. As turned out, the graphene deformation under external mechanical loading is extremely sensitive to the state of the edge atoms and makes it possible to disclose a topological nature of this sensitivity. Consider the example of the effect of uniaxial tension graphene that is a complicated mechano-chemical reaction [46, 47].

Figure 19 presents the (5, 5) NGr molecule, prepared for uniaxial tension in two orthogonal directions, that correspond to the zigzag (ZZ) and armchair (AC) modes of deformation. Red straight lines exhibit mechanochemical internal coordinates fixed by blue atoms in the pristine molecule. The deformation occurs as a stepwise elongation of these coordinates under fixed increment. Details of the reaction description are given elsewhere [46, 47].

The equilibrium structures of the (5, 5) NGr molecule before and after uniaxial tension, which was terminated by the rupture of the last C-C bond coupling two fragments of the molecule, are shown in Fig.20. Looking at the picture, two main peculiarities of the molecule deformation should be notified. First concerns the anisotropy of the deformation with respect to two deformational modes. Second exhibits a strong dependence of the deformation on the chemical composition of the molecule edge atoms.  As seen in the figure, the deformation behavior is the most complex for the naked molecule. The mechanical behavior of the (5, 5) NGr molecule is similar to that of a tricotage sheet when either the sheet rupture has both commenced and completed by the rupture of a single stitch row (AC mode) or rupture of one stitch is 'tugging at thread' the other stitches that are replaced by still elongated one-atom chain of the carbon atoms (ZZ mode). In the former case, the deformation is one-stage and is terminated on the 17[th] step of the deformation. In contrast, the ZZ mode is multi-stage and consists of 250 consequent steps with elongation of 0.1Å at each step [46, 47].

As turned out, the character of the deformation strongly depends on the chemical situation at the molecule edges. As seen in Fig.20b, the addition of one hydrogen atom to each of the molecule edge atoms does not change the general character of the deformation: it remains a tricotage-like one so that there is still a large difference between the behavior of ZZ and AC modes. At the same time, the number of the deformation steps in the first case reduces to 125.

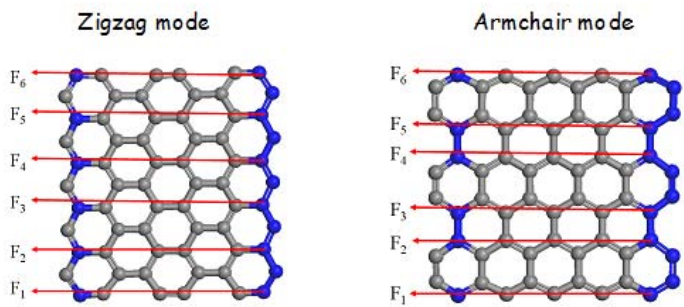

**Figure 19**. Six mechanochemical internal coordinates of uniaxial tension of the molecule (5,5) NGr for two deformation modes. $F_1$, $F_2$, $F_3$, $F_4$, $F_5$ и $F_6$ are forces of response along these coordinates. Blue atoms fix the coordinates starting lengths.

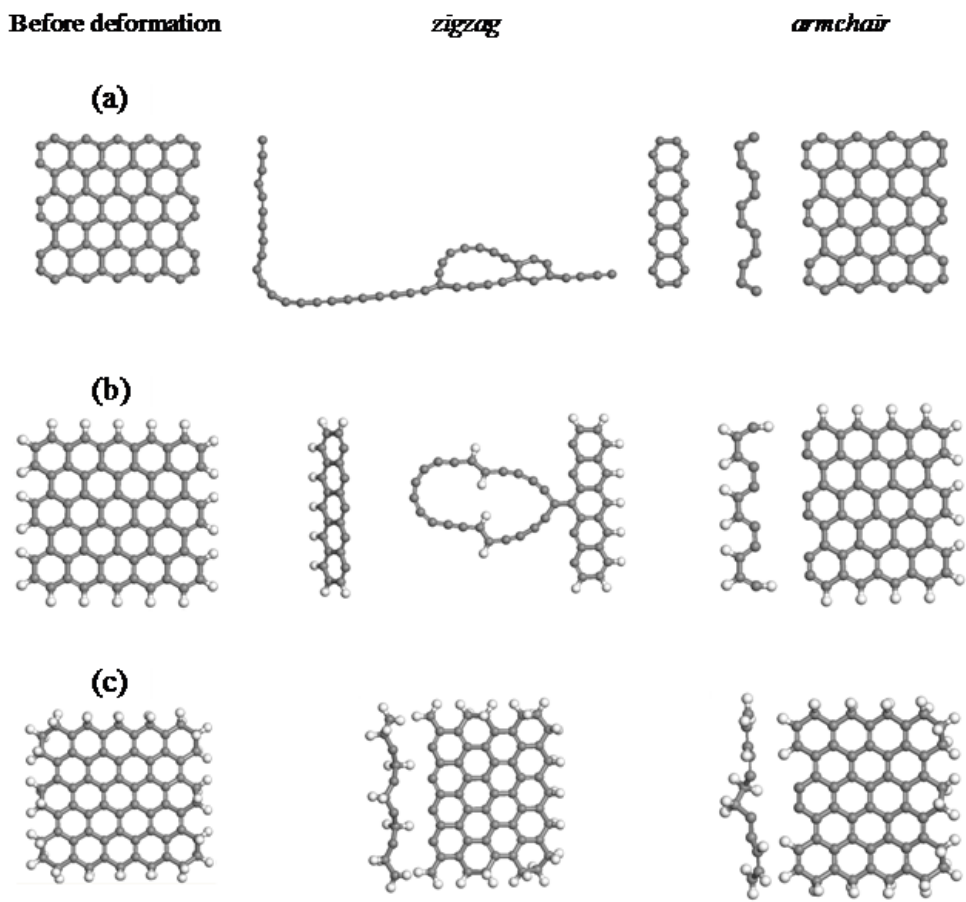

**Figure 20**. Equilibrium structures of the (5,5) NGr with different chemical modification of edge atoms before and after completing tensile deformation in two modes of deformation. Bare edges (a); $H_1$-terminated edges (b); $H_2$-terminated edges (c). White balls mark hydrogen atoms [45].

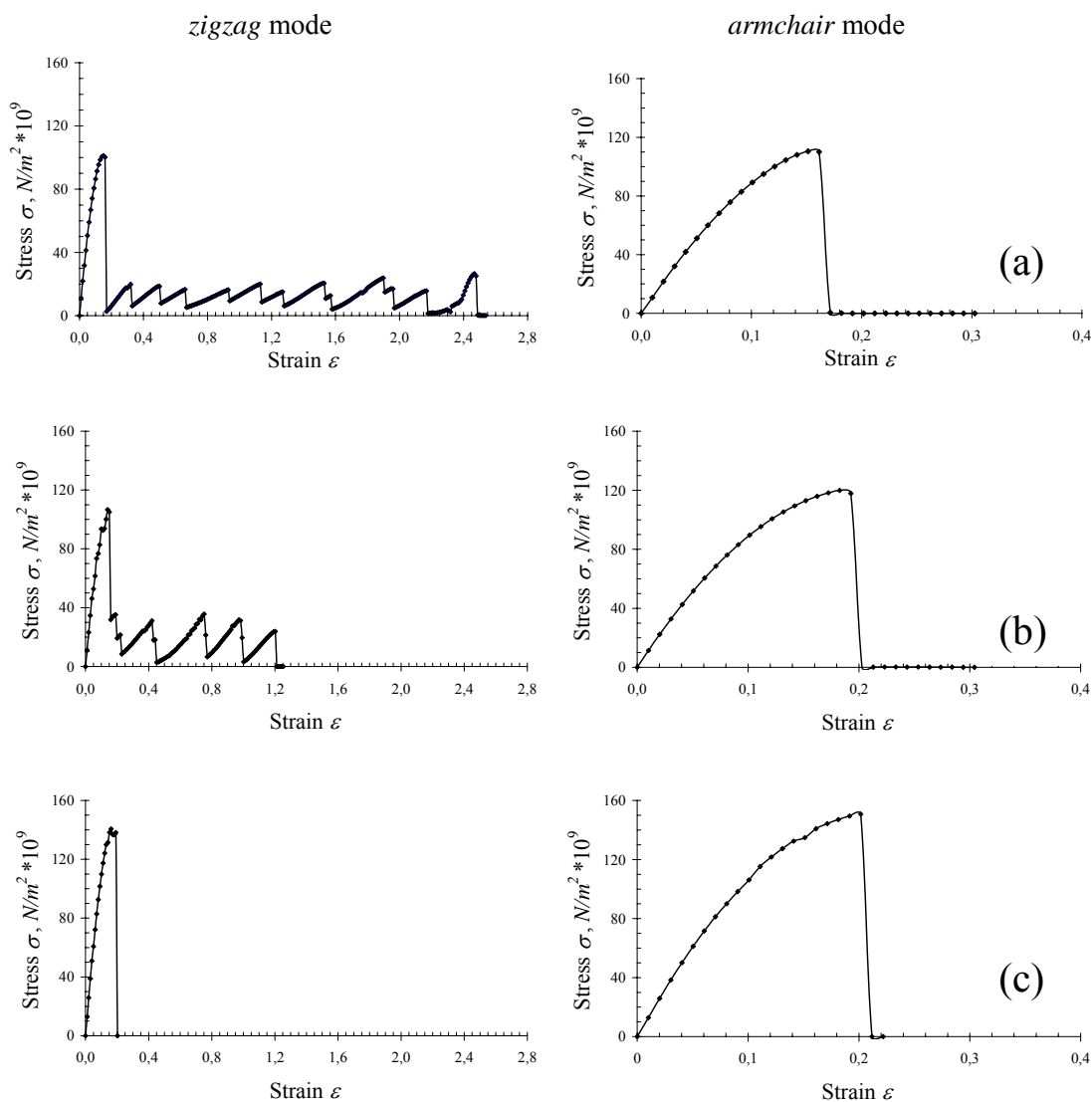

**Figure 21**. Stress-versus-strain dependences of tensile deformation of the (5, 5) NGr molecule with different chemical modification in two deformation modes. Nake molecule (a); $H_1$-terminated edges (b); $H_2$-terminated edges (c) [45].

Even more drastic changes for this mode are caused by the addition of the second hydrogen atoms to the edge ones (Fig.20c). Still, the AC mode is quite conservative while the ZZ one becomes practically identical to the former. The tricotage-like character of the deformation is completely lost, and the rupture occurs at the 20[th] step.

Figure 21 presents a set of the 'stress-strain' relations that fairly well highlight the difference in the mechanical behavior of all the three molecules. Table 4 presents the Young modules that were defined in the region of the elastic deformation. As seen from the table, the Young modules depend on the type of the edge atom chemical modification. As shown in [48], elastic properties of extended molecules such as polymers [49] and nanographenes [50] are determined by dynamic characteristics of the objects, namely, by force constants of the related vibrations. Since benzenoid units provide the determining resistance to any deformation of the graphene molecules, the dynamic parameters of the stretching C-C vibrations of the units are mainly responsible in the case of the uniaxial tension. Changing in Young's modules means changing in the force constants (and, consequently, frequencies) of these vibrations. The latter are attributed to the G-band of graphene that lays the foundation of a mandatory testing of any graphenium system by the Raman spectroscopy. In numerous cases, the relevant band is quite broad which might indicate that the chemical modification of the edge zone of the graphene objects under investigation may by one of the reasons of the band broadening.

**Table 4**. Young's modules for (5,5) NGr with different configuration of edge atoms, TPa

| Mode | Bare edges | $H_1$-terminated edges | $H_2$- terminated edges |
|---|---|---|---|
| zigzag | 1.05 | 1.09 | 0.92 |
| armchair | 1.06 | 1.15 | 0.95 |

Since the deformation-induced molecule distortion mainly concerns the basal atoms, so drastic changes in the deformation behavior points to a significant influence of the chemical state of the edge atoms on the electronic properties in the basal plane. The observed phenomenon can be understood if suggest that 1) the deformation and rupture of the molecule are a collective event that involves the electron system of the molecule as a whole; 2) the electron system of the graphene molecule is highly delocalized due to extreme correlation of the odd electrons; and 3) the electrons correlation is topologically sensitive due to which the chemical termination of the edge atoms strongly influences the behavior of the entire molecule.

## 6. Conclusive Remarks

A 'solid-and-molecule' dualism lays the foundation of the graphene uniqueness and chemical modification of graphene is the best platform for exhibiting its molecular essence. As follows from the molecular theory of graphene, due to rather long C-C chemical bonds, odd electrons of carbon atoms of graphene molecules are correlated and electrons with different spins are located in different sites of the space. The latter generates the molecule radicalization with a peculiar distribution of atomic chemical susceptibility (ACS) over the molecule atoms. The spatial ACS map exhibits a 'chemical portrait' of the molecule selecting atoms with the largest ACS as targets for addition reactions. The ACS map of naked graphene has a characteristic image showing that the graphene molecule of any size and shape is divided into two zones that cover edge and basal-plane carbon atoms, respectively, whose ACS values differ by 4 times in favor of the former while all the atoms of the molecule are chemically active. Due to the latter, first, the molecular chemistry of graphene is the chemistry of dangling bonds (as termed by Hoffmann [41]); second, any chemical modification of graphene molecules is a reaction that produces a number of polyderivatives; third, the graphene polyderivative structure can be convincingly predicted in the course of calculations following the ACS-pointer algorithms; fourth, the odd electrons correlation explains extremely strong influence of small changes in the molecule structure on properties as well as sharp response of the molecule behavior on small action of external factors. Taking together, the theory facilities have allowed for getting a clear, transparent and understandable explanation of hot points of the graphene chemistry and suggesting reliable models of the final products such as chemically produced and chemically reduced graphene oxides.

**Acknowledgement**

A financial support provided by the Ministry of Science and High Education of the Russian Federation grant 2.8223.2013 as well as of Russian Science Foundation grant 14-08-91376 is highly acknowledged.